\documentstyle{article}

\def\tilde{\widetilde}
\def\bar{\overline}
\def\hat{\widehat}
\def\*{\star}
\def\({\left(}          
\def\){\right)}         
\def\[{\left[}          
\def\]{\right]}

\def\proof{\noindent Proof. \hfill \break}
\def\frac#1#2{{#1 \over #2}}

\def\d{\partial}

\def\vev#1{\langle #1 \rangle}
\def\ket#1{ | #1 \rangle}
\def\bra#1{ \langle #1 |}

\def\2pi{\hbox{$2\pi i$}}

\def\dsl{\raise.15ex\hbox{/}\kern-.57em\partial}
\def\Dsl{\,\raise.15ex\hbox{/}\mkern-.13.5mu D}
         \def\Ga{\Gamma}
\def\b{\beta}           \def\hi{\chi }
\def\al{\alpha}

              \def\CC{{\cal C}}
\def\CD{{\cal D}}              
              
              \def\CL{{\cal L}}
\def\CM{{\cal M}}              \def\CO{{\cal O}}
\def\CP{{\cal P}}       \def\CQ{{\cal Q}}

       \def\CZ{{\cal Z}}

\def\debut{ \begin{eqnarray} }
\def\fin{ \end{eqnarray} }
\def\non{ \nonumber }

\def\w={={\kern -.55cm {\ }^{{\ }^{w}}}\ }

\def\square{\hfill
\vrule height6pt width6pt depth1pt \\}
\def\presentation{
\voffset -.50in
\hoffset -.19in
\oddsidemargin 0in \evensidemargin 0in
\marginparwidth .75in \marginparsep 7pt \topmargin 0in
\headheight 12pt \headsep .25in
\footheight 18pt \footskip .35in
\textheight 9.5in \textwidth 6.5in
\columnsep 10pt \columnseprule 0pt }

\presentation

\begin{document}
\rightline{SPhT/96/063}
\rightline{LPTHE-96-23}
\vskip 1cm
\centerline{\LARGE Null-vectors in Integrable Field Theory.}
\vskip 2cm
\centerline{\large O. Babelon ${}^{a\ 0}$, D. Bernard ${}^b$
\footnote[0]{Membre du CNRS} and F.A. Smirnov ${}^a$
\footnote[1]{On leave from Steklov Mathematical Institute,
Fontanka 27, St. Petersburg, 191011, Russia} }
\vskip1cm
\centerline{ ${}^a$ Laboratoire de Physique Th\'eorique et Hautes
Energies \footnote[2]{\it Laboratoire associ\'e au CNRS.}}
\centerline{ Universit\'e Pierre et Marie Curie, Tour 16 1$^{er}$
		\'etage, 4 place Jussieu}
\centerline{75252 Paris cedex 05-France}
\bigskip
 \centerline{ ${}^b$ Service de Physique Th\'eorique de Saclay
\footnote[3]{\it Laboratoire de la Direction des Sciences de la
Mati\`ere du Commissariat \`a l'Energie Atomique.}}
\centerline{F-91191, Gif-sur-Yvette, France.}
 \vskip2cm
{\bf Abstract.}
The form factor bootstrap approach allows to construct the space of
local fields in the massive restricted sine-Gordon model. This space
has to be isomorphic to that of the corresponding minimal model of conformal
field theory. We describe the subspaces which correspond to
the Verma modules of primary fields
in terms of the commutative algebra of local integrals of motion 
and of a fermion
(Neveu-Schwarz or Ramond depending on the particular primary field).
The description of null-vectors relies on the relation between
form factors and deformed hyper-elliptic integrals. The null-vectors
correspond to the deformed exact forms and to the deformed
Riemann bilinear identity. In the operator language, the null-vectors
are created by the action of two operators $\CQ$ (linear in the fermion) and
$\CC$ (quadratic in the fermion). We show that by factorizing out the
null-vectors one gets the space of operators with the correct character.
In the classical limit, using the operators $\CQ$ and $\CC$ we
obtain a new, very compact, description of the KdV hierarchy.
We also discuss a beautiful relation with the method of Whitham.
\newpage

\section{Introduction.}

In this article we present a  synthesis
of the ideas of the papers \cite{sm1}  and \cite{bbs}.
In the first of these papers the space of fields for the
sine-Gordon model (SG) was described in terms of the form factors
previously obtained in the bootstrap approach \cite{book}.
This description is based on rather special properties of
form factors for the SG model. Namely, it uses the fact that the form
factors were written in terms of deformed hyper-elliptic differentials,
allowing deformations of all the nice
properties of the usual hyper-elliptic differentials: the notion
of deformed exact forms and of the deformed Riemann bilinear identity
are available for them \cite{sm2}. Using these facts it has been
shown in \cite{sm1} that the same number of local
operators can be constructed in the generic case of the sine-Gordon model as
at the free fermion point.  The deformed exact forms and the deformed
Riemann bilinear identity are necessary in order to reduce the space
of fields to the proper size because in its original form factor description
the space is too big.
The description of \cite{sm1}
is basically independent of the coupling constant, but for
rational coupling constant there is a possibility to find additional
degenerations.

The problem with the description of the space of fields
obtained in \cite{sm1} is due to the fact that it is
difficult to compare it with the description coming from
Conformal Field Theory (CFT) or from the classical theory. The latter two are
closely connected because the Virasoro algebra can be
considered as a quantization of the second Poisson structure of KdV.
In the description of \cite{sm1}, it is even difficult to distinguish the descendents with
respect to the two chiral Virasoro algebras.

On the other hand in \cite{bbs} the semi-classical
limit of the form factor formulae has been understood.
This opens the possibility of identifying  all the
local operators by their classical analogues.
Using this result we decided to try to construct the
module of the descendents of the primary fields with respect
to the chiral Virasoro algebra. The result of this study happened
to be quite interesting.

Let us formulate more precisely the problems discussed in this paper.
The sine-Gordon model is described by the action:
$$ S={\pi\over\gamma}\int
\( (\partial _{\mu}\varphi)^2+
m^2 (\cos (2\varphi)-1)\)
\ d^2 x $$
where $\gamma$ is the coupling constant.
In the quantum theory, the relevant coupling constant is:
$\xi ={\pi \gamma\over \pi -\gamma} $.

The sine-Gordon theory
contains two subalgebras of local operators
which, as operator algebras, are generated by $\exp (i\varphi)$
and $\exp (-i\varphi)$ respectively. We shall consider one
of them, say the one generated by $\exp (i\varphi)$. It is known that
this subalgebra can be considered independently of the rest of the
operators, as the operator algebra of the theory with the
modified energy-momentum tensor:
\debut
T^{mod}_{\mu \nu}=T^{SG}_{\mu \nu}+i \al
\epsilon _{\mu,\mu '}\epsilon _{\nu,\nu '}
\partial _{\mu '}\partial _{\nu '}\varphi  \non
\fin
where $\al =\pi \sqrt {6\over \xi(\pi +\xi)} $.
This modification changes the trace of the
energy-momentum tensor which is now:
$T^{mod}_{\mu \mu}= m^2  \exp (2i\varphi) $.
This modified energy-momentum tensor corresponds
to the restricted sine-Gordon theory (RSG).
For rational ${\xi\over\pi}$, the RSG model describes
the $\Phi _{[1,3]}$-perturbations of the minimal models of CFT.
In this paper we consider only the RSG model.

It is natural from the physical point of view of integrable
perturbations \cite{zam} to expect that the space of fields for the perturbed
model is the same as for its conformal limit. The latter consists
of the primary fields and their descendents with respect to the
two chiral Virasoro algebras. In this paper we shall
consider the descendents with respect to one of these
algebras, the possibility of considering the descendents
with respect to the other one is
explained in Subsection 2.2. The Verma module of the
Virasoro algebra is generated by the action of the generators
$L_{-k}$ of the Virasoro algebra on the primary field.
The irreducible representation corresponding to a given
primary field is obtained by factorizing out the null-vectors
\cite{fefu,bpz}.

On the other hand the very possibility of integrable deformations
is due to the fact that there exists a commutative subalgebra
of the universal enveloping algebra of the Virasoro algebra: the algebra
of local integrals
\cite{zam,kuma,fefre}. The local integrals $I_{2k-1}$ have odd spins.
Logically it must be possible to present
the Verma module as a result of the action of the local
integrals of motion on a smaller module: the quotient over
the local integrals. The latter is isomorphic to the module created
from the vacuum by the action of bosons of even spins $J_{2k}$.
The important observation is that the form factor formulae allow us
to give a description of this space.
Certainly, such a realization of the Verma module is very
useful for integrable applications, but there is a difficulty.
It is not trivial to describe the null-vectors in this realization.
This question, of the description of the null-vectors, is the main problem
solved in this paper.

Briefly, the result is as follows.
It is useful to fermionize the bosons $J_{2k}$ by introducing
Neveu-Schwarz ($\psi _{2k-1}$) or Ramond ($\psi _{2k}$)
fermions, depending on which primary field we consider.
As in \cite{sm1} the null-vectors are due
to the deformed exact forms and the deformed Riemann bilinear
identity. In the fermionic language these null-vectors are created
by the action of two operators: a linear one in the fermion
($\CQ$) and a quadratic one ($\CC$).
By calculating the characters we show that, after factorizing
these null-vectors, we find exactly the same number of
descendents of a given primary field as in the corresponding
irreducible representation of the Virasoro algebra.

In Conformal Field Theory, the existence of null vectors provide
differential equations for the correlation functions. The null vectors
built out of $\CQ$ and $\CC$ in our approach will also lead to a system of
equations for the correlation functions in the massive case. These equations
will be similar to the hierarchies of equations in classical integrable
systems. In spite of the fact that such an infinite set of equations seems
{\it a priori} untractable, our hope is that, as in the classical  case,
large classes of solutions will be constructed. Hence we hope that these
equations will be really useful for the study of correlation functions
in the massive case.

We also present a detailed analysis of the classical limit of our
constructions.
The description of the null-vectors in terms of $\CQ$ and $\CC$
implies in the classical limit, a new and very compact description
of the KdV hierarchy.  This new description is not the same as the
description in terms of $\tau$-functions \cite{jap}, still the
technics we use are close to those of the Kyoto school.
Finally, we discuss an amazing analogy between the quantum
theory and the results of the Whitham method for small perturbations
around a given quasi-periodic classical solution of KdV.

Let us make one general remark on the exposition.
In this paper we restrict ourselves
to the reflectionless case of the sine-Gordon model. This is done only
in order to simplify the reading of the paper. All our
conclusions are valid for a  generic value of the coupling constant
as well.
For this reason, we chose to present the final formulae in the
situation of a generic coupling constant.
This leads, from time to time, to a contradictory situation.
 We hope that we shall be forgiven for this because
if we wrote generic formulae everywhere the
understanding of the paper would have been much more difficult.

\section{The space of fields.}

\subsection{The description of the space of fields in the $A,B$-variables.}

At the reflectionless points ($\xi={\pi\over\nu}$, $\nu=1,2,\cdots$)
there is a wide class of local operators $\CO$ for which
the form factors in the RSG-model
corresponding to a state with $n$-solitons and
$n$-anti-solitons are given by
\debut
&&f_\CO (\b _1,\b _2,\cdots ,\b _{2n})_{- \cdots -+\cdots +}=
\non \\
&&
=c^n\prod\limits _{i<j}\zeta (\b _i-\b _j)
\prod\limits _{i=1}^n \prod\limits _{j=n+1} ^{2n}
{1\over\sinh \nu (\b _j-\b _i -\pi i)}
\exp(-{1\over 2}(\nu (n-1)-n)\sum_j \b _j)
\non \\ && ~~~~~~~~~~~~~~~~~~\times
\ \widehat{f}_\CO (\b _1,\b _2,\cdots ,\b _{2n})_{-\cdots -+\cdots +}
\label{ff}
\fin
The function $\zeta (\b )$, without poles in the
strip $0< Im ~ \beta ~< 2\pi$, satisfies $\zeta (-\beta) =
S(\beta) \zeta (\beta)$
and $\zeta (\beta-2\pi i) =
\zeta (-\beta)$:
the S-matrix $S(\beta)$ and the constant
$c$ are given in the Appendix A.
The most essential part of the form factor is given by
\debut
\widehat{f}_\CO (\b _1,\b _2,\cdots ,\b _{2n})_{-\cdots -+\cdots +}&=&
\non\\
&&\hskip -5cm
={1\over (2\pi i)^n}\int dA_1\cdots \int dA_n
\prod\limits _{i=1}^n \prod\limits _{j=1} ^{2n} \psi (A_i,B_j)
\prod\limits _{i<j} (A_i^2-A_j^2)
\ L_\CO^{(n)} (A_1,\cdots ,A_n|B_1,\cdots ,B_{2n})
\prod\limits _{i=1}^n a_i^{-i} \label{ints}
\fin
where $B_j=e^{\b _j}$ and
\debut
 \psi (A,B)=\prod\limits _{j=1}^{\nu -1}(B -Aq^{-j} ),
\qquad {\rm with}~~ q= e^{i\pi/\nu}
\non
\fin
As usual we define $a=A^{2\nu}$.
Here and later if the range of integration is not specified the
integral is taken around 0. Notice that the operator dependence
of the form factors (\ref{ff}) only enters in $\widehat{f}_\CO$.

Different local operators ${\cal O}$ are defined by different
functions $L_\CO^{(n)} (A_1,\cdots ,A_n|B_1,\cdots ,B_{2n}) $.
These functions are symmetric polynomials of $A_1,\cdots ,A_n$.
For the primary operators $\Phi _{2k}=\exp (2k i\varphi )$
and their Virasoro descendants, $L_\CO$ are symmetric Laurent
polynomials of $ B_1,\cdots ,B_{2n} $. For the
primary operators $\Phi _{2k+1}=\exp ((2k+1) i\varphi)$,
they are symmetric Laurent polynomials of $ B_1,\cdots ,B_{2n} $
multiplied by $\prod B_j ^{1\over 2}$.
Our definition of the fields $\Phi _{m} $ is related to the notations coming
from CFT as follows: $\Phi _{m}$ corresponds to $ \Phi _{[1,m+1]}$.
The requirement of locality is guaranteed by the following simple
recurrent relation for the polynomials $L_{\CO}^{(n)}$:
\debut
&&\left . L_\CO^{(n)} (A_1,\cdots ,A_n|B_1,\cdots ,B_{2n})
\right|_{B_{2n}=-B_{1},\ A_n=\pm B_{1}}= \non\\
&&\hskip 1cm
=-\epsilon^\pm L_\CO^{(n-1)} (A_1,\cdots ,A_{n-1}|B_2,\cdots ,B_{2n-1})
\label{res}
\fin
where $\epsilon =+$ or $-$ respectively for the operators $\Phi _{2k} $
and their descendents, or for $\Phi _{2k+1} $ and their descendents.
In addition to the simple formula (\ref{ints}) we have
the requirement
\debut res _{A_n=\infty}\(
\prod\limits _{i=1}^n \prod\limits _{j=1} ^{2n} \psi (A_i,B_j)
\prod\limits _{i<j} (A_i^2-A_j^2)
\ L_\CO^{(n)} (A_1,\cdots ,A_n|B_1,\cdots ,B_{2n})a _n^{-k}\)=0,
\quad k\ge n+1
\label{vv}\fin
which is true in particular if
 $deg _{A_n}(L_{\CO})<2\nu$.
We explain this conditions in Appendix A. We see that 
the restriction (\ref{vv})
disappears only in the classical limit $\nu\to\infty$.
It will be clear later that this class of local operators is not 
complete
for the reason that the anzatz (\ref{ff}) is too restrictive.
With this anzatz we obtain the complete set of operators only in
the classical limit. However there is a possibility to
define the form factors of local operators which correspond to
polynomials satisfying the relation (\ref{res}) without any
restriction of the kind (\ref{vv}). To do that for the reflectionless
points one has to consider the coupling constant in generic
position (in which case the formulae for the form factors are
much more complicated \cite{book}) and to perform carefully the limit 
$\xi={\pi\over\nu}+\epsilon$, $\epsilon\to 0$.
An example of such calculation 
for $\xi =\pi$ is given in \cite{sm1}.
We shall give more comment about this procedure later, here
we would like to emphasize that the local operator
can be defined for any polynomial satisfying (\ref{res}) but
its form factors are not necessarily given by the anzatz (\ref{ff}).
Physically the existence of local operators for the
reflectionless case whose form factors are not given by the
anzatz (\ref{ff}) is related to existence of additional
local conserved quantities which constitute the algebra $\widehat{sl}(2)$.
In spite of the fact that the form factors of the form (\ref{ff})
do not define all the operators they provide a good example for
explaining the properties valid in generic case.

The explicit form of the polynomials $L_\CO$ for the primary operators
is as follows
$$ L_{\Phi _m}^{(n)}(A_1,\cdots ,A_n|B_1,\cdots ,B_{2n})
=\prod\limits _{i=1}^n A_i^m
\prod\limits _{j=1}^{2n} B_j^{-{m\over 2}} $$
In this paper we shall consider the Virasoro descendents of the
primary fields. We shall restrict ourselves by considering only
one chirality. Obviously, the locality relation (\ref{res}) is not
destroyed if we multiply the polynomial $L_\CO^{(n)}(A|B)$ either by
$I_{2k-1}(B)$ or by $J_{2k}(A|B)$ with
\debut
I_{2k-1}(B)&=&\({1+q^{2k-1}\over1-q^{2k-1}}\)
s_{2k-1}(B),\qquad k=1,2,\cdots\label{defI}\\
J_{2k}(A|B)&=&s_{2k}(A)- {1\over 2}s_{2k}(B),
\qquad k=1,2,\cdots
\label{defJ}
\fin
Here and later we shall use the following definition:
$$s_k(x_1,\cdots ,x_m)=\sum\limits _{j=1}^{m}x_j^{k}$$
The multiplication by $I_{2k-1}$ corresponds to the application of
the local integrals of motion. The normalization factor
${1+q^{2k-1}\over1-q^{2k-1}} $ is introduced for further
convenience. Since the boost operator acts by dilatation
on $A$ and $B$, $I_{2k-1}$ has spin $(2k-1)$ and $J_{2k}$ has spin $2k$.

The crucial assumption which we make is that the space
of local fields descendents of
the operator   $\Phi _{m} $ is generated by
the operators obtained from the generating function
\debut
\CL _m(t,y|A|B)=
\exp\Bigl(\sum\limits _{k\ge 1}t_{2k-1}I_{2k-1}(B)
+y_{2k}J_{2k}(A|B)\Bigr)~~
\({\prod\limits _{i=1}^n A_i^m
\prod\limits _{j=1}^{2n} B_j^{-{m\over 2}} }\) \label{gf1}
\fin
This is our main starting point.
This assumption follows from the classical meaning of the variables $A,B$
\cite{bbs}. We shall give classical arguments later, and
the connection with CFT is explained in the next subsection.

\subsection{The relation with CFT.}

The RSG model coincides with the $\Phi _{[1,3]}$-perturbation of
the minimal models of CFT. For the coupling constants which we
consider the minimal models in question are not unitary, but
this fact is of no importance for us.
It is natural from a physical  point of view,
to conjecture that the number of local operators
in the perturbed model is same as in its conformal limit.

We first need to recall a few basic facts concerning
the minimal conformal field theories.
At the coupling constant $\xi$, the conformal limit
of the RSG models have central charge
\debut
c = 1 - 6\frac{\pi^2}{\xi(\xi+\pi)} \non
\fin
The conformal limit of the fields $\Phi_m=e^{im\varphi}$ is
identified with the operators $\Phi_{[1,m+1]}$ in the minimal
models. They have conformal dimensions:
\debut
\Delta_m = \frac{m(m\xi - 2\pi)}{4(\xi+\pi)} \non
\fin
The reflectionless points, $\xi=\frac{\pi}{\nu}$ with
$\nu=1,2,\cdots$, correspond the non-unitary minimal
models $\CM_{(1,\nu+1)}$, which are degenerate cases.
At these points one has to
identify $\Phi_m$ with $\Phi_{2\nu-m}$.

The Virasoro Verma module corresponding to
the primary field $\Phi_{m}$ is generated by the vectors
$$L_{-k_1 }L_{-k_2 }\cdots   L_{-k_N }\Phi_{m} $$
At $\xi=\frac{\pi}{\nu}$ their structure
is the same as for generic value of $\xi$.
This means that for $m$ integer, the Verma module possesses
only one submodule. The so-called basic null-vectors, which we
shall denote by $\Ga_m$, are the generators
of these submodules. They appear at level $m+1$. The first few
are given by:
\debut
\Ga_0 &=& L_{-1}\Phi _0,\non\\
\Ga_1&=&(L_{-2}+\kappa L_{-1}^2)\Phi _1,
\qquad \kappa=-1-{\pi\over \xi}\non\\
\Ga _3&=&(L_{-3}+\kappa _1L_{-1}L_{-2}+\kappa _2L_{-1}^3)\Phi _2,
\quad \kappa _1=-2{\pi +\xi\over\pi+3\xi},
\kappa _2={(\pi +\xi)^2\over 2\xi(\pi +3\xi)}\non\\
\qquad &&etc~\cdots \non
\fin
Other null-vectors, whose set form the submodule,
 are created from the basic ones as follows
\debut
L_{-k_1 }L_{-k_2 }\cdots   L_{-k_N }~~ \Ga_m \non
\fin
As a consequence, the character of the irreducible Virasoro
representation with highest weight $\Phi_m$ is:
\debut
\chi_m(p)= \frac{1-p^{m+1}}{\prod_{j\geq1}(1-p^j)}
\non
\fin

\medskip
There exists an alternative description of these modules
which is more appropriate for our purposes.
Indeed, as is well known the integrability
of the $\Phi _{[1,3]}$-perturbation
is related to the existence of a certain commutative subalgebra of
the Virasoro universal enveloping algebra
generated by elements $I_{2k-1}$, polynomial of degree $k$
in $L_n$, such that $[L_0,I_{2k-1}]=(2k-1)I_{2k-1}$ and
\debut
[I_{2k-1},I_{2l-1} ]=0, \qquad k,l=1,2,\cdots \non
\fin
The first few are given by:
\debut
I_1&=& L_{-1},\non\\
I_3 &=& 2 \sum_{n\geq0}L_{-n-2}L_{n-1},\non\\
\qquad && etc~~\cdots\non
\fin
This is the subalgebra of the local integrals of motion.
The meaning of these operators and  of the $I_{2k-1}(B)$
which we had above is the same, so we denote them by the same
letters.

Of course acting only with this subalgebra on
$\Phi_m$ does not generate the whole Verma module.
But the left quotient of the Verma module by the ideal
generated by the integrals of motion produces
a space whose character is $\prod_{j\geq0}(1-p^{2j})^{-1}$.
This space can be thought of as generated by some
$J_{2k}$ with $[L_0,J_{2k}]=2kJ_{2k}$.

We expect that there is the following
alternative way of generating the Virasoro
module. Namely, in a spirit similar to the
Feigin-Fuchs construction \cite{fefu}, we expect that there exists
an appropriate completion of the subalgebra
of the integrals of motion by elements $J_{2k}$
such that (i) the Verma module
is isomorphic, as a graded space, to the space generated by the vectors
\debut
\exp(\sum\limits _{k\ge 1}t_{2k-1}I_{2k-1})
:\exp(\sum\limits _{k\ge 1}y_{2k}J_{2k}):\Phi_{m} \label{gf2}
\fin
where the double dots refer to an appropriate normal ordering,
and (ii)  that for certain normalization
of $I_{2k-1}$ and certain choice of $J_{2k}$  and their
normal ordering the generating
functions (\ref{gf1}) and (\ref{gf2}) provide different
realizations of the same object. We shall call the description
given by (\ref{gf1}) the $A,B$-representation of the Virasoro module.
The existence of this alternative description of the Virasoro
module is very important for integrable applications.


The main problem for this $I,J$ description arises from
the non-trivial construction of the null-vectors.
This problem has two aspects. First one has to
construct the null-vectors $\Ga_m$ in terms of $I,J$.
Then one also has to construct the submodule associated to it.
This is not as trivial as in the standard representation
since, while acting on $\Ga_m$ with the $I_{2k-1}$
still produces a null-vector, acting with the $J_{2k}$
does not necessarily leads to a null-vector.
We shall show that the whole null-vector submodule can be described
in the $A,B$-representation. The proof is based on rather
delicate properties of the form-factor integrals, the main of them being
the deformed Riemann bilinear identity \cite{sm2}.

\medskip
Let us discuss briefly the problem of the second chirality.
In \cite{bbs} we have explained that
the formula (\ref{ff}) has to be understood as a result
of light-cone quantization in which $x_-$ is considered
as space and $x_+$ as time. We must be able to consider the
alternative possibility ( $x_+$ - space, $x_-$ - time), and the results
of quantization must coincide. Where exactly the choice of
hamiltonian picture manifests itself in our formulae?
Consider the formula  (\ref{ints}). As it is explained in \cite{bbs}
the fact that $L_\CO^{(n)}$ is a polynomial in $A_i$ corresponds to
the choice of $x_-$ as space direction while the multiplier
$\prod\limits _{i=1}^n a_i^{-i}$ corresponds semi-classically to the choice
of trajectories under $x_+$-flow. These are the two ingredients
which change when the hamiltonian picture is changed.
Indeed, analyzing the results of \cite{bbs}, we find the following
alternative description of the form factors. The form factors are
given by the formulae (\ref{ff}) with $\widehat{f}_\CO$ replaced by
\debut
\widehat{h}_\CO (\b _1,\b _2,\cdots ,\b _{2n})_{-\cdots -+\cdots +}&=&
\non\\
&&\hskip -5cm
={1\over (2\pi i)^n}\int dA_1\cdots \int dA_n
\prod\limits _{i=1}^n \prod\limits _{j=1} ^{2n} \psi (A_i,B_j)
\prod\limits _{i<j} (A_i^2-A_j^2)
\ K_\CO^{(n)} (A_1,\cdots ,A_n|B_1,\cdots ,B_{2n})
\prod\limits _{i=1}^n a_i^{-i+1}\prod\limits_{j=1}^{2n} b_j^{-{1\over 2}}
\non
\fin
where $b_j=B_j^{2\nu}$,
$K_\CO^{(n)}$ is a polynomial in $A_i^{-1}$ and
the replacement of
$\prod a_i^{-i}$
by
$\prod a_i^{-i+1}\prod b_j^{-{1\over 2}}$
corresponds to the change of $x_+$-trajectories to  $x_-$-trajectories.
In particular for the primary fields we have
$$K_{\Phi _m}^{(n)}(A_1,\cdots ,A_n|B_1,\cdots ,B_{2n})
=\prod\limits _{i=1}^n A_i^{-m}
\prod\limits _{j=1}^{2n} B_j^{{m\over 2}} $$
The $\bar{L}_{-k}$ descendents are obtained by multiplying
$K_{\Phi _m}^{(n)}$ by $I_{-(2k-1)}(B)$ and $J_{-2k}(A|B)$.
The consistency of the two pictures requires that for the primary fields
they give the same result:
\debut
 \widehat{f}_{\Phi _m} (\b _1,\b _2,\cdots ,\b _{2n})_{-\cdots -+\cdots +}
=\widehat{h}_{\Phi _m} (\b _1,\b _2,\cdots ,\b _{2n})_{-\cdots -+\cdots +}
\label{chiralff}
\fin
This is a complicated identity which nevertheless can be proven.
We do not present the proof here because it goes beyond the scope of this paper.
It should be said, however, that the proof is based on the same technics
as used below (deformed Riemann bilinear identity {\it etc}).
Using the equivalence of the two representations of the form factors we can consider
the descendents with respect to $\bar{L}_{-k}$ and also mixed
$L_{-k},\bar{L}_{-k}$ descendents.
A formula similar to  (\ref{chiralff}) holds for any coupling constant.
However, at the reflectionless points there is another consequence of
(\ref{chiralff}): it also shows that the operators
$\Phi_m$ and $\Phi_{2\nu-m}$ are identified as it should be.

\section{Null-vectors.}

\subsection{The null-polynomials.}

Null-vectors correspond to operators with all the form factors
vanishing.
Consider the integral
\debut
{1\over (2\pi i)^n}\int dA_1\cdots \int  dA_n
\prod\limits _{i=1}^n \prod\limits _{j=1} ^{2n} \psi (A_i,B_j)
\prod\limits _{i<j} (A_i^2-A_j^2)
\ L_\CO^{(n)} (A_1,\cdots ,A_n|B_1,\cdots ,B_{2n})
\prod\limits _{i=1}^n a_i^{-i} \label{inte}
\fin
Instead of $L_\CO^{(n)}$, we shall often use the anti-symmetric
polynomials $M_\CO^{(n)} $:
\debut
&&M_\CO^{(n)} (A_1,\cdots ,A_n|B_1,\cdots ,B_{2n})=
\prod\limits _{i<j} (A_i^2-A_j^2)
\ L_\CO^{(n)} (A_1,\cdots ,A_n|B_1,\cdots ,B_{2n})
\non \fin
The dependence
on $B_1,\cdots ,B_{2n}$ in the polynomials  $M_\CO^{(n)} $ will
often be omitted.

There are several reasons why this integral can vanish. Some of
them depend on a particular value of the coupling constant or
on a particular number of solitons. We should not consider these
occasional situations. There are three general reasons for
the vanishing of the integral, let us present them.

{\bf 1. Residue.} The integral (\ref{inte}) vanishes if vanishes the residue
with respect to $A_n$ at the point $A_n=\infty$
of the expression
$$
\prod\limits _{j=1} ^{2n} \psi (A_n,B_j)a_n^{-n}
M_\CO^{(n)} (A_1,\cdots ,A_n)
$$
Of course the distinction of the variable $A_n$ is of no importance
because $M_\CO^{(n)} (A_1,\cdots ,A_n)$ is anti-symmetric.

{\bf 2. "Exact forms."}
The integral (\ref{inte}) vanishes if
$M_\CO^{(n)} (A_1,\cdots ,A_n)$ happens to be
an "exact form". Namely, if it can be written as:
\debut
&&M_\CO^{(n)} (A_1,\cdots ,A_n) =\sum\limits _k (-1)^k
M(A_1,\cdots ,\widehat{A_k},\cdots ,A_n)
\(Q(A_k)P(A_k)-qQ(qA_k)P(-A_k)\),
\label{zero}
\fin
with
\debut
P(A)=\prod_{j=1}^{2n} (B_j+A) \non
\fin
for some anti-symmetric polynomial $M(A_1,\cdots ,A_{n-1})$.
Here and later $\widehat{A_k}$ means that $A_k$ is omitted.
This is a direct consequence of the functional equation satisfied by
$\psi(A,B)$ (See eq.(\ref{fonc}) in Appendix B).
For $Q(A)$ one can take in principle any Laurent polynomial,
but since we want $M_\CO^{(n)}$ to be a polynomial
the degree of $Q(A)$ has to be greater or equal $-1$.

{\bf 3. Deformed Riemann bilinear relation.} The integral
(\ref{inte}) vanishes if
\debut
M_\CO^{(n)} (A_1,\cdots ,A_n)=
\sum\limits _{i<j}(-1)^{i+j}M(A_1,\cdots ,\widehat{A_i},
\cdots ,\widehat{A_j},\cdots A_n)C(A_i,A_j)
\non \fin
where $M(A_1,\cdots ,A_{n-2})$ is an anti-symmetric polynomial
of $n-2$ variables, and $C(A_1,A_2)$ is given by
\debut
C(A_1,A_2 )={1\over A_1A_2}\left\{ {A_1-A_2\over A_1+A_2 }
(P(A_1)P(A_2)-P(-A_1)P(-A_2))
+
(P(-A_1)P(A_2)-P(A_1)P(-A_2))\right\}  \label{C}
\fin
This property needs some comments. For the case of generic coupling constant
its proof is rather complicated. It is a consequence of the so
called deformed Riemann bilinear
identity \cite{sm2}.
The name is due to the fact that
in the limit $\xi\to\infty$ the deformed
Riemann bilinear identity happens to be
the same as the Riemann bilinear identity for hyper-elliptic integrals.
The formula for $C(A_1,A_2)$ given in \cite{sm3} differs from (\ref{C})
by simple "exact forms". Notice that the formula for $C(A_1,A_2)$
does not depend on the coupling constant.
For the reflectionless case a
very simple proof is available which is given in Appendix B.

In order to apply these restrictions to the description of the null-vectors
we have to make some preparations. The expression for $C(A_1,A_2)$ given above
is economic in a sense that, as a polynomial of $A_1,A_2$ it has
degree $2n-1$, but it is not appropriate for our goals because it
mixes odd and even degrees of $A_1,A_2$ while the descendents
of primary fields contain only odd or only even polynomials.
So, by adding an "exact forms" we want
to replace $C(A_1,A_2)$ by equivalent expressions
of higher degrees which contain only odd or only even degrees.

\proclaim
Proposition 1.
The following equivalent forms of $C(A_1,A_2)$ exist:
\debut
C(A_1,A_2)\simeq C_e(A_1,A_2) \simeq C_o(A_1,A_2) ,
\label{CCC}
\fin
here and later $\simeq$ means equivalence up to "exact forms".
The formulae for  $C_e(A_1,A_2)$ and $ C_o(A_1,A_2)$ are as follows:
\debut
&&C_e(A_1,A_2)=
\int\limits _{{|D_2|>|D_1| \atop |D_1|>|A_1|,|A_2|} }
dD_2 dD_1 ~
\tau _e \left({D_1\over D_2}\right)
\left( {P(D_1)P(D_2)\over(D_1^2-A_1^2)(D_2^2-A_2^2)}-
{P(D_1)P(D_2)\over(D_2^2-A_1^2)(D_1^2-A_2^2)} \right) \label{CNS}
\fin
where
$$\tau _e(x)=\sum\limits _{k=1}^{\infty}{1-q^{2k-1}\over 1+q^{2k-1}}x^{2k-1} -
\sum\limits _{k=1}^{\infty}{1+q^{2k}\over 1-q^{2k}}x^{2k} $$
and
\debut
&&C_o(A_1,A_2)=A_1A_2
\int\limits _{{|D_2|>|D_1| \atop |D_1|>|A_1|,|A_2|}} dD_2dD_1
\tau _o \left({D_1\over D_2}\right)\
\left({P(D_1)P(D_2)\over(D_1^2-A_1^2)(D_2^2-A_2^2)}-
{P(D_1)P(D_2)\over(D_2^2-A_1^2)(D_1^2-A_2^2)} \right)
\non \fin
where
$$\tau _o(x)=\sum\limits _{k=1}^{\infty}{1-q^{2k}\over 1+q^{2k}}x^{2k} -
\sum\limits _{k=1}^{\infty}{1+q^{2k-1}\over 1-q^{2k-1}}x^{2k-1} $$
\par

The proof of this proposition is given in Appendix B.

The functions $\tau _e(x)$ and $\tau _o(x)$ are not well defined when
$q^r=1$ because certain denominators vanish. However the formula
(\ref{CCC}) in which the LHS is independent of $q$ implies that
in the dangerous places we always find "exact forms" . So, for our
applications these singularities are harmless. We shall comment more on this 
point later.

\subsection{The fermionization.}

The descendents of the local operators are created by $I_{2k-1}$ and $J_{2k}$.
This generates a bosonic Fock space.
It is very convenient to fermionize $J_{2k}$. Let us introduce Neveu-Schwarz
and Ramond fermions: $\psi _{2k-1}, \psi _{2k-1}^{*}$ and
$\psi _{2k}, \psi _{2k}^{*}$.
The commutation relations are as follows
$$ \psi _{l}\psi _{m}^{*}+ \psi _{m}^{*} \psi _{l} =\delta _{l,m}$$
We prefer to follow the notations from \cite{jap} than those
coming from CFT, the reader used to CFT language has to replace $\psi ^*_m$
by $\psi ^*_{-m}$.

The vacuum vectors
for the spaces with different charges are defined as follows.
In the Neveu-Schwarz sector we have:
\debut
\psi _{2k-1}|2m-1\rangle =0,~~ for\ k>m,\qquad
\psi _{2k-1}^{*} |2m-1\rangle =0,~~ for\ k\le m ; \non\\
\langle 2m-1 |\psi _{2k-1}=0,~~ for\ k\le m,\qquad
\langle 2m-1 |\psi ^{*}_{2k-1}=0,~~ for\ k > m\non
\fin
For the Ramond sector we have:
\debut
\psi _{2k}|2m\rangle =0,~~ for\ k>m,\qquad
\psi _{2k}^{*} |2m\rangle =0,~~ for\ k\le m ; \non\\
\langle 2m |\psi _{2k}=0,~~ for\ k\le m\qquad
\langle 2m |\psi ^{*}_{2k}=0,~~ for\ k >m  \non
\fin
We shall never mix the Neveu-Schwarz and Ramond sectors. The spaces spanned by
the right action of an equal number of $\psi$'s and $\psi ^{*}$'s on
the vector $\langle p |$ will be called $H_p^*$.
The right action of $\psi$ sends $H_p^*$ to $H_{p+2}^*$.
It is useful to think of the vector $\langle p |$ as a semi-infinite product
$$ \langle p | = \cdots \psi_{p-4}\psi_{p-2}\psi_{p} $$

Let us introduce generating functions for the fermions.
The operators $\psi(A),\psi ^{*}(A)$ are defined for the Neveu-Schwarz and
the Ramond sectors respectively as
follows
\debut
\psi(A)&=\sum\limits _{k=-\infty}^{\infty}A^{-2k+1}\psi _{2k-1},\qquad
\psi^{*}(A)=\sum\limits _{k=-\infty}^{\infty}A^{2k-1}\psi ^{*}_{2k-1};
\non\\
\psi(A)&=\sum\limits _{k=-\infty}^{\infty}A^{-2k}\psi _{2k},\qquad
\psi^{*}(A)=\sum\limits _{k=-\infty}^{\infty}A^{2k}\psi ^{*}_{2k}
\non
\fin
We shall use the decomposition of $\psi(A),\psi ^{*}(A)$ into
the regular and singular parts (at zero):
$$\psi(A)=\psi(A)_{reg} +\psi(A)_{sing},\qquad
\psi ^{*}(A)=\psi ^{*}(A)_{reg} +\psi ^{*}(A)_{sing}$$
where $ \psi(A)_{reg}$ and $ \psi ^*(A)_{reg}$ contain
all the terms with non-negative degrees of $A$.

Let us introduce the bosonic commuting operators $h_{-2k}$ for $k\ge 1$:
\debut
&&h_{-2k}= \sum\limits _{j=-\infty}^{\infty}
\psi _{2j-1}\psi _{2k+2j-1}^{*}\qquad for\ Neveu-Schwarz\ sector,\non\\
&&h_{-2k}= \sum\limits _{j=-\infty}^{\infty}
\psi _{2j}\psi _{2k+2j}^{*}\qquad for\ Ramond\ sector,\non
\fin
They satisfy the commutation relations:
$[h_{-2k},h^*_{-2l}]= -k \delta_{k,l}$.
We also have the following  commutation relations between
the fermions and the bosons:
\debut
\psi (A)h_{-2k}=(h_{-2k}-A^{-2k} )\psi (A)
\label{comm}
\fin

The bosonic
generating function $\CL _m(t,y|A|B)$ for the descendents of the operators
$\Phi _{m}$ can be rewritten as:
\debut
{\CL}_m(t,y|A|B)=
\exp\({\sum\limits_{k \ge 1}t_{2k-1} I_{2k-1}(B)}\)~~
\bra{m-1} \exp\bigl(\sum_{k\ge 1} y_{2k}h^*_{-2k}\bigr)~
\widehat{\CL}_m(A|B) \ket{m-1}
\label{xxbis}
\fin
where
\debut
\widehat{\CL}_m(A|B)=
\exp\({-\sum\limits_{k \ge 1}{1\over k}h_{-2k} J_{2k}(A,B)}\)~~
\prod_i A_i^m\prod_j B_j^{-{m\over 2}} \non
\fin
In other words,
to have a particular descendent one has to take in the expression
\debut
\exp\({\sum\limits_{k \ge 1}t_{2k-1} I_{2k-1}(B)}\)
\ \widehat{\CL}_m(A|B) \ket{m-1}
\non
\fin
the coefficient in front of some monomial in $t_{2k-1}$
and to calculate the matrix element with some vector from
the fermionic space $H^*_{m-1}$.

We can replace this bosonic expression by a fermionic one.
Recall that as a direct result of the boson-fermion correspondence
one has:
$$\prod\limits _{i<j}(A_i^2-A_j^2)
\exp\(-\sum\limits_{k \ge 1}{1\over k}h_{-2k}
\sum\limits _{i=1}^nA_i^{2k}\)|m-1\rangle =
\psi^{*}(A_1)\cdots\psi^{*}(A_n)|m-2n-1\rangle \prod_{i=1}^n A_i^{-m+2n-1}
$$
where the fermions are Neveu-Schwarz or Ramond ones depending on the parity
of $m$.  Using this fact
the formula (\ref{xxbis}) can be rewritten for any $n$ as follows
\debut
\prod\limits _{i<j}(A_i^2-A_j^2)~~ \widehat{\CL}_m(A|B) |m-1\rangle=
\ g(B) ~\ \psi^{*}(A_1)\cdots\psi^{*}(A_n)|m-2n-1\rangle~~
\prod_i A^{2n-1}_i\prod_j B_j^{-{m\over 2}}
\non
\fin
where
\debut
g(B)=\exp\({\sum\limits_{k \ge 1}{{1\over 2k}}h_{-2k} s_{2k}(B)}\)
\label{gB}
\fin
The occurrence of this operator is similar to that of the
spin field in the description \cite{kni} of CFT on
hyperelliptic curves.

Let us now concentrate on the operator $\Phi _0=1$ and its descendents.
It means that we are working with the $H_{-1}^*$
subspace of the Neveu-Schwarz sector.
The polynomials corresponding to the operators in this sector are
even in $A_i$.
Let us describe the null-vectors due to the restrictions 1, 2, 3,
from the previous subsection.
To do that we shall need the results of the following three propositions.

\proclaim Proposition 2.
The set of polynomials of the form
\debut
M_\CO^{(n)} (A_1,\cdots ,A_n)=
\sum\limits _{i<j}(-1)^{i+j}M(A_1,\cdots ,\widehat{A_i},
\cdots ,\widehat{A_j},\cdots A_n)C _e(A_i,A_j) \label{zz}
\fin
coincides up to "exact forms" with the set of the matrix elements:
\debut
\langle \Psi _{-5} |\widehat{\CC }
\ \psi^{*}(A_1)\cdots\psi^{*}(A_n)|-2n-1\rangle \prod A^{2n-1}_i,
\qquad \forall \ \Psi _{-5}\in H^*_{-5} \label{zeroC^}
\fin
where
\debut
\widehat{\CC}=
\int\limits _{|D_2|>|D_1|}dD_2 \int\limits dD_1
P(D_1)P(D_2)D_1^{-2n-1}D_2^{-2n-1}\tau _e \left({D_1\over D_2}\right)
\psi(D_1)\psi(D_2) \non
\fin
\par
\proof
The vectors from the space $H^*$ can be written as
$$\langle \Psi |=
\bra{N}\psi _{k_1}\cdots\psi _{k_p}
$$
where $k_p>\cdots >k_1>N+1$. We shall call $N$ the depth of $\langle \Psi |$.
There are three possibilities
for the matrix element (\ref{zeroC^}) to differ from zero:

1. The depth of $\langle \Psi _{-5}|$ is greater than $-2n-1$

2. The vector $\langle \Psi _{-5}|$
is obtained from a vector $\langle \Psi _{-1}|$
whose depth is greater than $-2n-1$ by application

of $\psi ^{*}_{-2p-1} \psi ^{*}_{-2q-1}$
with $q>p\ge n$ (i.e. there are two holes below $-2n-1$).

3. The vector $\langle \Psi _{-5}|$
is obtained from a vector $\langle \Psi _{-3}|$
whose depth is greater than $-2n-1$ by application

of $\psi ^{*}_{-2p-1}$
with $p\ge n$ (i.e. there is one hole below $-2n-1$).  \newline
In the first case using the formula
$$\langle -2n-1|\psi (D) \psi ^{*}(A)|-2n-1\rangle =
\left({D\over A}\right)^{2n-1}{D^2\over (D^2-A^2)},\qquad |D|>|A| $$
and (\ref{CNS}) one find
\debut
&&\langle \Psi _{-5} |
\widehat{\CC}
\ \psi^{*}(A_1)\cdots\psi^{*}(A_n)|-2n-1\rangle
\prod A^{2n-1}_i =\non\\
&&\hskip 1cm =\sum\limits _{i<j}(-1)^{i+j}M(A_1,\cdots ,\widehat{A_i},
\cdots ,\widehat{A_j},\cdots A_n)C _e(A_i,A_j) \non
\fin
where
$$
M(A_1,\cdots ,A_{n-2})=
\langle \Psi _{-5} |
\psi^{*}(A_1)\cdots\psi^{*}(A_{n-2})|-2n-1\rangle
\prod A^{2n-1}_i
$$
In the second case it is necessary that in the expression
$ \langle \Psi _{-1} | \psi ^{*}_{-2p-1}
\psi ^{*}_{-2q-1} \ \widehat{\CC}$
the two holes below $-2n-1$
are annihilated by $ \widehat{\CC} $, the result is
$$ \langle \Psi _{-1} |
\int\limits _{|D_2|>|D_1|}dD_2 \int\limits dD_1
P(D_1)P(D_2)D_1^{-2n+2p}D_2^{-2n+2q}\tau _e \left({D_1\over D_2}\right)=0
$$
because the integrand is a regular function of $D_1$
for $p\ge n$.

In the third case it is necessary that in the expression
$ \langle \Psi _{-3} | \psi ^{*}_{-2p-1}
\ \widehat{\CC}$
the hole below $-2n-1$ is annihilated by $ \widehat{\CC} $, the result is
$$ \langle \Psi _{-3} |
\int\limits _{|D_2|>|D_1|}dD_2 \int\limits dD_1
P(D_1)P(D_2)D_1^{-2n-1}D_2^{-2n+2p}\tau _e
\left({D_1\over D_2}\right)\psi(D_1)
$$
where the pairing of $\psi (D_1)$ and $\psi ^{*}_{-2p-1}$ is
not considered because it produces zero for the same reason as above.
In the matrix element we shall have the polynomials
\debut
&&\langle -2n-1|\int\limits _{|D_2|>|D_1|}dD_2 \int\limits dD_1
P(D_1)P(D_2)D_1^{-2n-1}D_2^{-2n+2p}\tau _e
\left({D_1\over D_2}\right)\psi(D_1)\psi ^{*}(A_j)|-2n-1\rangle A_j^{2n-1}=
\non\\&&=
\int\limits _{|D_2|>|D_1|}dD_2 \int\limits dD_1
P(D_1)P(D_2){1\over D_1^2-A_j^2}D_2^{-2n+2p}\tau _e
\left({D_1\over D_2}\right)\equiv R_{2n+2p}(A_j)
\non\fin
The polynomial $R_{2n+2p}(A)$ is an even polynomial of degree
$2n+2p$. Let us show that it is an "exact form". From the calculations
of Appendix B we have
\debut
&&R_{2n+2p}(A)=
\int\limits _{|D_2|>|D_1|}dD_2 \int\limits dD_1
P(D_1)P(D_2){1\over D_1^2-A^2}D_2^{-2n+2p}\tau _e
\left({D_1\over D_2}\right)\simeq\non\\
&&\simeq A\int\limits _{|D_2|>|D_1|}dD_2 \int\limits dD_1
P(D_1)P(D_2){1\over D_1^2-A^2}D_2^{-2n+2p}
\left({1\over D_1+D_2}\right)=\non\\&&=
A\int\limits _{|D_1|>|D_2|}dD_2 \int\limits dD_1
P(D_1)P(D_2){1\over D_1^2-A^2}D_2^{-2n+2p}
\left({1\over D_1+D_2}\right)=0
\label{pol}\fin
where we have changed the integral over $|D_2|>|D_1|$
by the integral over $|D_1|>|D_2|$ because the residue
at $D_2=-D_1$ gives the integral over $D_1$ of an even
function, such integral equals zero. The last integral in
(\ref{pol}) vanishes because the integrand is a regular
function of $D_2$ for $p\ge n$.
Let us emphasize that our construction is self-consistent
because for every $n$ the polynomials of too high degree
(greater than $4n-2$) are "exact forms".

Thus we have non-trivial matrix elements only in the first case
which
obviously exhausts the polynomial of the kind (\ref{zz}). \square

Consider now the restriction 2
of the previous subsection. It is easy to figure out that
there is only one uniform way to write for all $n$
polynomials of the type (\ref{zero})
which are even in all variables $A_i$. Namely:
\debut
&&M_\CO^{(n)} (A_1,\cdots ,A_n) =
\sum\limits _k (-1)^k
M(A_1^2,\cdots ,\widehat{A_k^2},\cdots ,A_n^2) \left(P(A_k)-
P(-A_k)\right)A_k^{-1}
\label{zzz}
\fin
where $M(A_1^2,\cdots ,\cdots ,A_{n-1}^2)$  is arbitrary anti-symmetric
polynomial.

The following two simple propositions are given without proof.

\proclaim
Proposition 3. The set of polynomials (\ref{zzz})
coincides with the set of matrix elements:
\debut
\langle \Psi _{-3} |\widehat{\CQ}
\ \psi^{*}(A_1)\cdots\psi^{*}(A_n)|-2n-1\rangle \prod A^{2n-1}_i
\qquad \forall \Psi _{-3}\in H^*_{-3}\non
\fin
where
\debut
\widehat{\CQ}=\int dD D^{-2n-1} P(D)\psi (D) \non
\fin
\par

\proclaim
Proposition 4.
The set of polynomials $M_\CO^{(n)} (A_1,\cdots ,A_n)$
such that
$$res _{A_n=\infty}
\prod_j\psi (A_n,B_j)a_n^{-n} M_\CO^{(n)} (A_1,\cdots ,A_n) =0$$
(we hope that the same letter $\psi$ used for the function $\psi (A,B_j)$
and for the fermion is not confusing)
coincides with the set of matrix elements:
\debut
\langle \Psi _{1} |\widehat{\CQ} ^{\dag}
\psi^{*}(A_1)\cdots\psi^{*}(A_n)|-2n-1\rangle
\prod A^{2n-1}_i \qquad \forall
\Psi _{1}\in H^*_{1} \non
\fin
where
\debut
\widehat{\CQ} ^{\dag} =
res _{A=\infty} \prod_j\psi (A,B_j)a^{-n}
\int\limits _{|D|>|A|} \psi ^*(D){1\over D^2-A^2}D^{2n}dD
\non \fin
\par


Let us apply these results to the description of null-vectors.
We need to introduce the following notations:
\debut
&&X(D)=
\sum\limits _{k\ge 1}{1\over 2k-1}D^{-2k+1}s_{2k-1}(B)=
\sum\limits _{k\ge 1}D^{-2k+1}
{1\over 2k-1}\({1-q^{2k-1}\over 1+q^{2k-1}}\)I_{2k-1}(B),
\label{defXD}
\\
&& Y(D)=
\sum\limits _{k\ge 1}{1\over 2k}D^{-2k}s_{2k}(B)\label{defYD}
\fin
Obviously
\debut
P(D)=D^{2n}e^{X(D)-Y(D)} \label{PenXY}
\fin
The null-vectors will be produced by acting with
some operators $\CC$, $\CQ$ and $\CQ^{\dag}$ on $\widehat{\CL}_0(A|B)|-1\rangle$.
In view of the bosonization formulae, these operators
are obtained from $\widehat{\CC}$, $\widehat{\CQ}$ and
$\widehat{\CQ}^{\dag}$ by conjugation with $g(B)$:
\debut
\CC \ g(B)=g(B)\ \widehat{\CC},\qquad
\CQ \ g(B)=g(B)\ \widehat{\CQ},\qquad
\CQ^{\dag} \ g(B)=g(B)\ \widehat{\CQ}^{\dag} \non
\fin
The formulae for $\CC$ and $\CQ$ are
given in the following  two propositions:

\proclaim
Proposition 2'.
 From the Proposition 2 we find that the null-vectors
due to the deformed Riemann bilinear identity are of the form:
\debut
\exp(\sum\limits_{k \ge 1}t_{2k-1} I_{2k-1}(B))~~
\langle \Psi _{-5} |\ \CC~
\exp(-\sum\limits_{k \ge 1}{{1\over k}}h_{-2k} J_{2k}(A,B)) |-1\rangle
\non \fin
where
\debut
\CC=
\int\limits _{|D_2|>|D_1|}{dD_2\over D_2} \int\limits {dD_1\over D_1}
e^{X(D_1)}e^{X(D_2)}\tau _e \left({D_1\over D_2}\right)
\psi(D_1)\psi(D_2) \label{CNSF}
\fin
\par

\proclaim
Proposition 3'.
 From the Proposition 3 one gets the following set of null-vectors
due to the "exact forms":
\debut
\exp(\sum\limits_{k \ge 1}t_{2k-1} I_{2k-1}(B))~~
\langle \Psi _{-3} |\CQ
\exp(-\sum\limits_{k \ge 1}{{1\over k}}h_{-2k} J_{2k}(A,B)) |-1\rangle
\non \fin
where
\debut
\CQ= \int\limits {dD\over D} e^{X(D)} \psi(D) \label{defcq1}
\fin
\par
Notice a very important feature in these formulae: The operators $\CQ$ and
$\CC$ are independent of $n$.

These propositions are direct consequences of the previous
ones and of the following conjugation property of the fermions:
\debut
\psi(D)~g(B) = g(B)~\psi(D)~ e^{-Y(D)}, \qquad
\psi^*(D)~g(B) = g(B)~\psi^*(D)~ e^{Y(D)} \non
\fin

Before dealing with $\CQ ^{\dag}$
let us discuss the operator $\CC$, $\CQ$
 in more details. It will be convenient to rewrite them in terms
of another set of fermions $\tilde \psi$ and $\tilde \psi^{\dag}$.
To understand the purpose of introducing a new basis
for the fermions consider the formula
(\ref{CNSF}). In this formula the fermion $\psi (D_2)$
can be replaced by its regular part
$\psi(D_2)_{reg}$
because other multipliers in the integrand contain only negative
powers of $D_2$. That is why $\CC$ can be rewritten in the form
\debut
\CC=
\int\limits {dD\over D}
\tilde{\psi}(D)_{sing}\tilde{\psi}(D)_{reg}    \label{zzzz}
\fin
where the modified fermion $\tilde{\psi}$ is defined as follows
\debut
\tilde{\psi}(D)_{reg}= \psi(D)_{reg} ,\qquad
\tilde{\psi}(D)_{sing}=U\psi\ (D),
\non\fin
with $U$ the following operator
\debut
Uf\ (D)=
\left[\int\limits _{|D|>|D_1|}{dD_1\over D_1}
e^{X(D_1)}e^{X(D)}\tau _e \left({D_1\over D}\right)
f(D_1)\right]_{odd}
\non \fin
where $[\cdots]_{odd}$ means that only odd degrees of the expression
with respect to $D$ are taken because only those contribute to  the
integral (\ref{zzzz}).
It is quite obvious that this transformation is triangular, namely
$$ \tilde{\psi}_{2k-1}=\({1-q^{2k-1}\over 1+q^{2k-1}}\)
\psi_{2k-1} +(terms\ with\  \psi_{2l-1} ,l<k),
\qquad k\ge 1$$
Altogether we can write $\tilde{\psi}(D)=\hat{U}\psi\ (D)$ where
$\hat{U}$ is triangular.

Introduce the fermions $\tilde{\psi}^{\dag} $
satisfying canonical commutation relations with $\tilde{\psi} $:
\debut
\tilde{\psi}^{\dag}(D)=\(\hat{U}^{-1}\)^T\psi^{*}\ (D)
\non\fin
Since the operator $\hat{U}$ is not unitary, we do not use
$*$ but $\dag$ for $\tilde{\psi}$.
The triangularity of the operator $\hat{U}$ guaranties that
the Fock space $H^*$ constructed in terms of $\tilde{\psi}, \tilde{\psi}^{\dag}$
coincides with the original one.

Thus, we can rewrite (\ref{zzzz}) as follows
\debut
\CC=\sum\limits _{j=1}^{\infty}
\tilde{\psi} _{-2j+1}  \tilde{\psi}_{2j-1}  \non
\fin
The important property of this formula is that for given number
of solitons $n$ the summation can be taken from $1$ to $n$ because
the operators $\tilde{\psi}_{2j-1} $ with $j>n$ produce "exact forms"
when plugged into the matrix elements (see the proof of Proposition 2).

Similarly, we can express the operator $\CQ$, defined
in (\ref{defcq1}), in terms
of $\tilde \psi$:
\debut
\CQ = \int\limits {dD\over D} e^{X(D)} \tilde{\psi}(D)
\non
\fin
This equality is due to the fact that only the regular
part of $\psi(D)$ contributes into the integral which does not
change under the transformation to $\tilde{\psi}(D)$.

Now we are ready to consider the operator $\CQ^{\dag}$.

\proclaim Proposition 4'.
 From Proposition 4 one gets the following
set of null-vectors due the vanishing of the residues:
\debut
\exp(\sum\limits_{k \ge 1}t_{2k-1} I_{2k-1}(B))~~
\langle \Psi _{1} |\CQ ^{\dag}
\exp(-\sum\limits_{k \ge 1}{{1\over k}}h_{-2k} J_{2k}(A,B)) |-1\rangle
\non \fin
where
\debut
\CQ ^{\dag}=\int\limits {dD\over D} e^{X(D)} \tilde{\psi}^{\dag}(D)
\non\fin
\par
\proof
Directly from Proposition 4 one gets the following formula
for $\CQ ^{\dag}$:
\debut
\CQ ^{\dag}
= res _{A=\infty} \prod\limits _j\psi (A,B_j)a^{-n}
\int\limits _{|D|>|A|}dD D^{2n}
e^{-Y(D)} \psi^{*}(D) {1\over D^2-A^2}
\label{ppp}
\fin
This formula looks much simpler in terms of $\tilde{\psi}^{\dag}$.
By definition we have
$$ \psi ^{*}(D) =\widehat{ U}^T\tilde{\psi}^{\dag}\ (D)=
\left[\int\limits _{|D_1|>|D|}{dD_1\over D_1}
e^{X(D_1)}e^{X(D)}\tau _e \left({D\over D_1}\right)
\tilde{\psi}^{\dag}_{reg}(D_1)     \right]_{odd}+
\tilde{\psi}^{\dag}(D)_{sing} $$
The last term does not contribute to the residue because
$$\int\limits _{|D|>|A|}dD D^{2n}
e^{-Y(D)} {1\over D^2-A^2}
 \tilde{\psi}^{\dag}(D)_{sing}
=\CO(A^{2n-2})$$ 
and
\debut
\prod_j \psi (A,B_j)a^{-n} =A^{-2n}(1+\CO(A^{-1}))\label{cat}\fin
Substituting the rest into (\ref{ppp}) one has
\debut
&&\CQ ^{\dag}= \int {dD_1\over D_1}\tilde{\psi}^{\dag}(D_1)e^{X(D_1)}
res _{A=\infty}\prod\limits _j \psi (A,B_j)a^{-n}
\int\limits_{|D_1|>|D|} dD P(D)
 {1\over D^2-A^2} \tau _e \left({D\over D_1}\right)
\non
\fin
Using the formulae from Appendix B one can show that
\debut
\int\limits_{|D_1|>|D|>|A|} dD P(D)
 {1\over D^2-A^2} \tau _e \left({D\over D_1}\right) \simeq
{1\over 2}\({P(A)\over A+D_1}+{P(A)\over A-D_1}\)+\CO(A^{-1})=
A^{2n-1}(1+ \CO(A^{-1}))\non\fin
Here the equality is up to "exact form" in $A$; such "exact form" never contribute
to the residue.  Now the formula (\ref{cat}) gives
$$res _{A=\infty}\prod_j \psi (A,B_j)a^{-n}
\int\limits_{|D_1|>|D|>|A|} dD P(D)
 {1\over D^2-A^2} \tau _e \left({D\over D_1}\right)=1$$
which proves the proposition. \square

This alternative expression for $\CQ^{\dag}$ shows that it is independent of
$n$, as $\CQ$ and $\CC$ are.

Notice that in the formulae for $\CQ$ and $\CQ ^{\dag}$
only the holomorphic parts of $\tilde{\psi}(D)$ and $\tilde{\psi}^{\dag}(D)$
are relevant. This  leads to the 
important commutation relation:
\debut
[\CC,\CQ ^{\dag}]=\CQ  \label{comm1}
\fin
Notice also that $\CQ$ and $\CQ ^{\dag}$ are nilpotent operators, 
$\CQ ^2=(\CQ ^{\dag})^2=0$, and $[\CC,\CQ]=0$.

This is the proper place to discuss the problems which we had before:
the definition of local operator corresponding to arbitrary
polynomial satisfying (\ref{res}) and the singularities
in the definition of $C_o$ and $C_e$.

Consider the polynomials
\debut
\prod\limits _{i<j}(A_i^2-A_i^2)\langle \tilde{\Psi} _{-3} |\CQ
\exp(-\sum\limits_{k \ge 1}{{1\over k}}h_{-2k} J_{2k}(A,B)) |-1\rangle
\label{co} \fin 
where the states $\langle \tilde{\Psi} _{-3} |$ are the vectors
from the Fock space constructed via the fermions the $\tilde{\psi}$.
By the very definition of $\tilde{\psi}$ the polynomial (\ref{co})
is equivalent to  an antisymmetric polynomial
$M_{\langle \tilde{\Psi} _{-3} |}(A_1,\cdots ,A_n)$ of degree
$\le 2n-1$ with respect to any $A_i$ which is
{\it independent of the coupling constant}. This polynomial
satisfies the requirement formulated in \cite{sm1} and hence
defines a local operator for arbitrary coupling constant.
So, if we start from the fermions $\tilde{\psi}$ the counting of
the local operators is independent of the coupling constant like
in the paper \cite{sm1}. But where is the origin of vanishing
denominators? The point is that if we consider
the rational coupling constant $\xi$, for many
of these local operators all the form factors will vanish.
So, we have to be more careful: to consider the vicinity of
the rational coupling constant $\xi =\pi{p\over q}+\epsilon$ and to
keep the first non-vanishing order in the form factors in the
limit $\epsilon\to 0$. Thus, to define the space of local operators 
it is better to start from the fermions $\tilde{\psi}$.
If when passing to the fermions $\psi$ we find 
somewhere infinite coefficient when the coupling
constant is rational, it is always accompanied by a vanishing
operator in such a way that the result is finite.

\subsection{Algebraic definition of the null-vectors.}

Let us summarize our study of the null-vectors for the descendents of 
$\Phi _0$.
It is quite convenient to present the null-vectors as vectors from the
dual space. In this section
we shall write $\langle \Psi |\ \w= 0$ if the matrix element
of $\langle \Psi |$ with the fermionic generating function
of local fields vanishes under the integral ($w$ means "in a weak sense").
We have found three types of null-vectors:
\debut
\matrix{\langle \Psi _{-5}|\CC\ \w= 0,\qquad
&\forall\quad \langle\Psi _{-5}|  \in H^*_{-5},&\qquad (i)\non\\
\langle \Psi _{-3}|\CQ\ \w= 0,\qquad
&\forall\quad \langle\Psi _{-3}|  \in H^*_{-3},&\qquad (ii)\non\\
\langle \Psi _1|\CQ^{\dag}\ \w= 0,\qquad
&\forall\quad \langle\Psi _1|  \in H^*_{1},&\qquad (iii)\non
}
\fin
where the operators $\CC$, $\CQ ^{\dag}$, $\CQ$ are given
respectively by
\debut
\CC=
\int\limits {dD\over D}
\tilde{\psi}(D)_{reg}\tilde{\psi}(D)_{sing},
\qquad \CQ =\int {dD\over D} e^{X(D)} \tilde{\psi}(D),
\qquad \CQ ^{\dag}=\int {dD\over D}e^{X(D)} \tilde{\psi}^{\dag}(D)
\non
\fin
Let us show that these three conditions for the null-vectors are not
independent.

It is easy to show that the operator $\CC$ identifies the
spaces $H_{-3}^*$ and $H_{1}^*$:
$$ Ker(\CC |_{H_{-3}^*\to H_{1}^*})=0,\qquad
Im(\CC |_{H_{-3}^*\to H_{1}^*})=H_1^* $$
Hence every $\langle \Psi _1|  \in H^*_{1}$ can be presented as
$ \langle \Psi _{-3}|\CC$ for some $\langle \Psi _{-3}|\in H_{-3}^*$.
Let us show that the null-vectors $(iii)$ are linear
combinations of $(i)$ and $(ii)$. We  have:
\debut
\langle \Psi _1|\CQ^{\dag}=\langle \Psi _{-3}|\CC\CQ^{\dag}=
\langle \Psi _{-3}|\CQ+\langle \Psi _{-3}|\CQ^{\dag}\CC
\non \fin
where we have used the commutation relation (\ref{comm1}).
Thus we have proven the following
\proclaim Proposition 5.
In the space of descendents of
$\Phi _0=1$ which is $H(I)\otimes  H^*_{-1} $ (where $H(I)$
is the space of polynomials of $\{I_{2k-1}\}$)
the null-vectors coincide with the vectors
\debut
\matrix{\langle \Psi _{-5}|\CC\ \w= 0,\qquad
&\forall\quad \langle\Psi _{-5}|  \in H^*_{-5},&\qquad (i)\non\\
\langle \Psi _{-3}|\CQ\ \w= 0,\qquad
&\forall\quad \langle\Psi _{-3}|  \in H^*_{-3},&\qquad (ii)\non
}
\fin
and their descendents with respect to $I$'s.
\par

The consideration of the other operators $\Phi _{2m}$ is based on the same
formulae, but involves additional complications. We do not want to go
into details, and we only present the final result.
\proclaim Proposition 6.
For the operator
$\Phi _{2m}$
whose descendents are counted by
the vectors from  $H(I)\otimes H^*_{2m-1}$
we have two types of independent null-vectors:
\debut
\langle \Psi _{-5-2m}|(\CC)^{m+1}\ \w= 0,\qquad
\forall\quad \langle\Psi _{-5-2m}|  \in H^*_{-5-2m},\qquad &(i)\non\\
\langle \Psi _{-3-2m}|(\CC)^{m}\CQ\ \w= 0,\qquad
\forall\quad \langle\Psi _{-3-2m}|  \in H^*_{-3-2m},\qquad &(ii)\non
\fin
and their descendents with respect to $I$'s.
\par
For the operators $\Phi _{2m+1}$ one has the following picture.
There is no uniform ``exact form'' which is an odd polynomial, so,
the analogue of the operator $\CQ$ does not exist for $\Phi _{2m+1}$, and
the null-vectors are either due to the deformed
Riemann bilinear identity or due to the vanishing
of the residue.
To construct the null-vectors in terms of fermions one
has to introduce first the operator $\CC$:
\debut
\CC=
\int\limits _{|D_2|>|D_1|}{dD_2\over D_2} \int\limits {dD_1\over D_1}
e^{X(D_1)}e^{X(D_2)}\tau _o \left({D_1\over D_2}\right)
\psi(D_1)\psi(D_2)
\non \fin
where the fermions are from the Ramond sector.
This operators can be rewritten in a form similar to  (\ref{zzzz}):
\debut
\CC=
\int\limits {dD\over D}
\tilde{\psi}(D)_{reg}\tilde{\psi}(D)_{sing}  \non
\fin
The fermions $\tilde{\psi} $ are related to  $\psi$ by triangular
transformation.  

The consideration of the operator $\Phi _1$ is absolutely parallel
to the consideration of $\Phi _0$. The null-vectors are
created either by the action of $\CC$ (Riemann identity) or
by the action of $\tilde{\psi}^{\dag}_0$ (residue).
Notice that
$$ [\CC,\tilde{\psi}^{\dag}_0]=0 $$
which guaranties the consistency.
For higher operators $\Phi _{2m+1}$ there are additional problems which
we would not like to discuss here. The general result is given
in the following
\proclaim Proposition 7.
The descendents of the operator $\Phi _{2m+1}$ are counted by the
vectors from the space  $H(I)\otimes H^*_{2m}$.
We have two types of null-vectors
\debut\matrix{
\langle \Psi _{-4-2m}|(\CC)^{m+1}\ \w= 0,\qquad
&\forall\quad \langle\Psi _{-4-2m}|  \in H^*_{-4-2m},\qquad &(i)\non\\
\langle \Psi _{2-2m}|(\CC)^{m}\tilde{\psi}^{\dag}_0 \ \w= 0,\qquad
&\forall\quad \langle\Psi _{2-2m}|  \in H^*_{2-2m},\qquad &(ii)\non }
\fin
and their descendents with respect to $I$'s
\par

\subsection{Examples of null-vectors and the characters.}

Let us present the simplest examples of null-vectors for the
operators $\Phi _0$, $\Phi_1$ and $\Phi _2$.

For the operator $\Phi _0$ the simplest null-vector is created by
$$\langle -3|\CQ=s_1(B)\langle -1|$$
This null-vector is
$$\({1-q\over 1+q}\)I_1 \Phi _0 $$
This null-vector is to be compared with
$$L_{-1} \Phi _0 $$

For the operator $\Phi _1$ the simplest null-vector is
created by
$$\langle 2|\tilde{\psi}^{\dag}_0=
\langle -2|\tilde{\psi} _2=
\({1-q^2\over 1+q^2}\)
\left(\langle -2|\psi _{2} - {1\over 2}
\({1+q\over 1-q}\)^2 s_1(B)^2\langle 0|\right)
$$
which gives the null-vector
$$\({1-q^2\over 1+q^2}\)(J_2- {1\over 2}I_1^2)\Phi _1$$
which has to be compared with
$$  (L_{-2}+\kappa L_{-1}^2)\Phi _1$$
For the operator $\Phi _2$ the simplest null-vector 
is created by $\langle -5|\CC \CQ$. It yields
$$
{1\over 3}\({1-q^3\over 1+q^3}\)
(I_3 - 3 I_1J_2 +{1\over 2} I_1^3)\Phi _2
$$
which has to be compared with
$$ (L_{-3}+\kappa _1L_{-1}L_{-2}+\kappa _2L_{-1}^3)\Phi _2$$
Notice that the relative coefficients in our parametrization
of the  null-vectors are
independent of $\xi$. So, they are exactly of the same form as
the classical one. However, this is not always the case.

Let us show that generally the number of our null-vectors is
the same as for the representations of the Virasoro algebra.
Recall that we consider the null-vectors which do not depend
on the arithmetical properties of ${\xi\over\pi}$, so, there is one
basic null-vector in every Verma module of Virasoro algebra. The
character of the irreducible module associated with $\Phi _m$ is
\debut
\hi _m(p)=(1-p^{m+1}) {1\over\prod _{j\ge 1}(1-p^j)}
\label{cVir}
\fin
where we omitted the multiplier with the scaling dimension
of the primary field. We can not control this scaling dimension,
the dimensions of the descendents are understood relatively to
the dimension of the primary field.
The character (\ref{cVir}) is obtained from the character
of the Verma module by omitting the module of descendents
of the null-vector on the level $m+1$.

Let us consider the character of the module which we constructed
in terms of $I,J$. The dimensions of $I_{2k-1}$ and $J_{2k}$ are
naturally $2k-1$ and $2k$. If we do not take into account the null-vectors,
the characters of all the modules associated with $\Phi _m$ are the
same:
$$\hi(p) = {1\over\prod _{j\ge 1}(1-p^j)} $$
Let us take into account the null-vectors. They are described
in terms of fermions. By consistency with the dimensions
of $I_{2k-1}$ and $J_{2k}$ one find that the dimensions of
$\psi _l$ and $\psi _{-l}^{\dag}$ equal $l$.

Technically it is easier
to start with $\Phi _{2m+1}$.
The space of descendents is $H(I)\otimes  H^*_{2m} $ where $H(I)$
is the space of polynomials of $\{I_{2k-1}\}$.
\proclaim Proposition 8.
The character of the space of descendents of $\Phi_{2m+1}$, modulo the null
vectors, equals
 $$ \hi _{2m+1}(p)=(1-p^{2(m+1)}){1\over \prod _{j\ge 1}(1-p^{j})} $$
\par
\proof
The null-vectors are defined in the Proposition 7.
It is easy to eliminate the null-vectors $(ii)$: we have to consider the
subspace $ H_{-2m,\ 0}^*$ in which the level $\tilde{\psi}_0$ is
always occupied.
Consider the sequence
\debut
&&H^*_{-2m-4,\ 0}
\ {\rightarrow{\kern -.5cm {{\ }^{\CC}}}\ }
H^*_{-2m ,\ 0}
\ {\rightarrow{\kern -.5cm {{\ }^{\CC ^m}}}\ }
H^*_{2m,\ 0}\non
\fin
The operator  $\CC^{m}$ identifies the spaces
$H_{-2m,\ 0}^*$ and $H_{2m,\ 0}^*$:
$$ Ker(\CC ^m |_{H_{-2m,\ 0}^*\to H_{2m,\ 0}^*})=0,
\qquad Im(\CC ^m |_{H_{-2m,\ 0}^*\to H_{2m,\ 0}^*})=H_{2m,\ 0}^* $$
and $\langle -2m-2|\tilde{\psi}^{\dag}_0\CC^{m} =  \langle 2m|  $.
Hence we can count the descendents by vectors of the space
$ H(I)\otimes H_{-2m,\ 0}^* $ with the null-vectors:
\debut
\langle \Psi _{-4-2m}|\CC\ \w=0,\qquad
&\forall\quad \langle\Psi _{-4-2m}|  \in H^*_{-4-2m, \ 0},\non
\fin

Notice that the operator $\CC$ is dimensionless and
$$Ker(\CC |_{H_{-4-2m,\ 0}^*\to H_{-2m,\ 0}^*})=0 $$
That is why for the character of the space of descendents without
null-vectors we have:
\debut
\hi _{2m+1}(p)={1\over \prod _{j\ge 1}(1-p^{2j-1})}p^{-m(m-1)}
\left(\hi _{{\ }_{H_{-2m,\ 0}^*}}(p)
-\hi _{{\ }_{H_{-2m-4,\ 0}^*}}(p) \right)  \label{gg}
\fin
where the first multiplier comes from $H(I)$ the multiplier
$p^{-m(m-1)}$ is needed in order to cancel the dimension
of the vacuum vector in $ H_{-2m}^*$. Let us evaluate the
expression in brackets:
\debut
&&\hi _{{\ }_{H^*_{-2m,\ 0}}}(p)-\hi _{{\ }_{H_{-2m-4,\ 0}^*}}(p) =\non\\
&&=\int \prod\limits _{j\ge 1}(1+p^{2j}x)(1+p^{2j}x^{-1}) x^{-m}{dx\over x}-
\int \prod\limits _{j\ge 1}(1+p^{2j}x)(1+p^{2j}x^{-1}) x^{-m-2}{dx\over x}=
\non\\
&&=\int (1+x^{-1})\prod\limits _{j\ge 1}(1+p^{2j}x)(1+p^{2j}x^{-1})
x^{-m}{dx\over x}  -
\int (1+x^{-1})\prod\limits _{j\ge 1}(1+p^{2j}x)(1+p^{2j}x^{-1})
x^{-m-1}{dx\over x} =\non\\
&&=(1-p^{2(m+1)})\int (1+x^{-1})\prod\limits _{j\ge 1}
(1+p^{2j}x)(1+p^{2j}x^{-1}) x^{-m}{dx\over x} =
\non\\&&
=(1-p^{2(m+1)})\hi_{{\ }_{H_{-2m}^*}}(p)
=p^{m(m-1)}(1-p^{2(m+1)}) {1\over \prod _{j\ge 1}
(1-p^{2j})} \non
\fin
where we have changed the variable of integration $x\to
xp^{-2}$ in the second integral when passing from third to forth line.
Substituting this result into (\ref{gg}) we get the correct character:
$$ \hi_{2m+1}(p)=(1-p^{2(m+1)}){1\over \prod _{j\ge 1}(1-p^{j})} $$
\square

Let us consider now the operators $\Phi _{2m}$.
We parametrize the descendents of $\Phi _{2m}$ by
the vectors from  $H(I)\otimes H^*_{2m-1}$,
\proclaim Proposition 9.
The character of the space of descendents of $\Phi_{2m}$, modulo the
null-vectors, equals
 $$ \hi _{2m}(p)=(1-p^{2m+1}){1\over \prod _{j\ge 1}(1-p^{j})} $$
\par
\proof
The null-vectors are defined in the Proposition 6. We have
\debut
&&H^*_{-2m-5}
\ {\rightarrow{\kern -.5cm {{\ }^{\CC}}}\ }
H^*_{-2m-1 }
\ {\rightarrow{\kern -.5cm {{\ }^{\CC ^m}}}\ }
H^*_{2m-1}\non\\
&&H^*_{-2m-3}
\ {\rightarrow{\kern -.5cm {{\ }^{\CQ}}}\ }
H^*_{-2m-1}
\ {\rightarrow{\kern -.5cm {{\ }^{\CC ^m}}}\ }
H^*_{2m-1}\non
\fin
The operator $\CC ^m$ identifies $H^*_{-2m-1}$ and $H^*_{2m-1}$:
$$ Ker(\CC ^m |_{H_{-2m-1}^*\to H_{2m-1}^*})=0,
\qquad Im(\CC ^m |_{H_{-2m-1}^*\to H_{2m-1}^*})=H_{2m-1}^* $$
and $\langle -2m-1|\ \CC^{m} =  \langle 2m-1|  $.
Hence we can replace the space
$H(I)\otimes H^*_{2m-1}$ with these null-vectors
by $H(I)\otimes H^*_{-2m-1}$
with null-vectors
\debut \matrix{
\langle \Psi _{-5-2m}|\CC\ \w=0,\qquad
&\forall\quad \langle\Psi _{-5-2m}|  \in H^*_{-5-2m},\qquad &(i)\non\\
\langle \Psi _{-3-2m}|\CQ\ \w=0,\qquad
&\forall\quad \langle\Psi _{-3-2m}|  \in H^*_{-3-2m},\qquad &(ii)\non}
\fin
So, the character in question is
\debut
\hi _{2m}(p)={1\over \prod _{j\ge 1}(1-p^{2j-1})}p^{-m^2}
\left(\hi _{{\ }_{H^*_{-2m-1,\ 0} }}(p)-
\hi _{{\ }_{H^*_{-2m-5,0}}}(p)\right)
\non
\fin
where $H^*_{-2l-1,\ 0}=H^*_{-2l-1}/H^*_{-2l-3}\CQ $.
In order to calculate the character $\hi _{{\ }_{H^*_{-2l-1,\ 0}} }(p)$
one has to take into account that $\CQ $ is a nilpotent operator, $\CQ ^2=0$ ,
with a trivial cohomology. Hence
$$ Ker(\CQ |_{H_{-2j-3}^*\to H_{-2j-1}^*})=
Im(\CQ |_{H_{-2j-5}^*\to H_{-2j-3}^*}) $$
Summing up over this complex we obtain:
$$ \hi _{{\ }_{H^*_{-2l-1,\ 0} }}(p)=
\int\limits _{|x|>1}
\prod\limits _{j\ge 1}(1+p^{2j-1}x)(1+p^{2j-1}x^{-1}) x^{-l}
{x\over x+1}{dx\over x} $$
Hence
\debut
&&\hi _{{\ }_{H^*_{-2m-1,\ 0} }}(p)-
\hi _{{\ }_{H^*_{-2m-5,\ 0}}}(p)=\non\\
&&=\int\limits _{|x|>1}
\prod\limits _{j\ge 1}(1+p^{2j-1}x)(1+p^{2j-1}x^{-1}) x^{-m}
{x\over x+1}{x^2-1\over x^2}{dx\over x} =\non\\
&&=\int \prod\limits _{j\ge 1}(1+p^{2j-1}x)(1+p^{2j-1}x^{-1}) x^{-m}
{dx\over x} -
\int \prod\limits _{j\ge 1}(1+p^{2j-1}x)(1+p^{2j-1}x^{-1}) x^{-m-1}
{dx\over x} =\non\\
&&=(1-p^{2m+1})\hi _{{\ }_{H^*_{-2m-1}}}(p) =
p^{m^2} (1-p^{2m+1})  {1\over \prod _{j\ge 1}(1-p^{2j})}
\non
\fin
Thus the character is given by
$$\hi _{2m}(p)=(1-p^{2m+1}){1\over \prod _{j\ge 1}(1-p^{j})} $$
as it should be. \square

\section{Classical case.}
\subsection{Local fields and null-vectors in the classical theory.}

The classical limit of the light-cone component $T_{--}$
 of the energy-momentum
tensor gives the KdV field $u(x _-)$.
When working
with the multi-time formalism we shall identify $x_-$ with $t_1$.
Local fields in the KdV theory, descendents of the identity operators, are
simply polynomials in $u(t)$ and its derivatives  with respect to $t_1$:
\begin{eqnarray}
\CO &=& \CO (u,u',u'',...)
\label{loc1}
\end{eqnarray}
We shall use both notations $\partial _1$ and $'$ for the
derivatives with respect to $x_- =t_1$.

Instead of the variables $u,u',u'',...$, it will be more convenient to replace
 the odd derivatives of $u$ by the higher time derivatives $\partial_{2k-1} u$,
 according to the equations of motion
\begin{eqnarray}
{\partial L \over \partial t_{2k-1}}
&=& \left[ \left( L^{2k-1\over 2} \right)_+, L
\right] = {1 \over 2^{2k-1}}u^{(2k-1)}+ \cdots   \nonumber
\end{eqnarray}
Here
$$L = \partial_1^2 - u$$
is the Lax operator of KdV. We have used the pseudo-differential
operator formalism.  We follow the book \cite{dickey}.

The even derivatives of
$u(x)$ will be replaced by the densities $S_{2k}$ of the local integrals of
motion,
\begin{eqnarray}
S_{2k} &=& res_{\partial _1} L^{2k-1\over 2} =-{1\over 2^{2k-1}} u^{(2k-2)}+
\cdots,  \nonumber \end{eqnarray}
In particular on level 2 we have $S_2=-{1\over 2}u$.
For a reader who prefers the $\tau$-function language $S_{2k}=\partial_1
\partial _{2k-1} \log \tau$.

 From analogy with the conformal case we put forward the following
main conjecture underlying the classical picture:

\proclaim Conjecture.
We can write any local fields as
\begin{eqnarray}
{\cal O} (u,u',u'',...) &=& F_{{\cal O},0} (S_2, S_4, \cdots) +
\sum_{\nu \geq 1}\partial^\nu
F_{{\cal O},\nu} (S_2, S_4, \cdots)
\label{loc2}
\end{eqnarray}
where $\nu =( i_1,i_3, \cdots)$ is a multi index,
 $\partial^\nu = \partial^{i_1}_{1} \partial^{i_3}_{3} \cdots $,
 $|\nu | = i_1 + 3 i_3 +\cdots$.

\par

We have checked this conjecture up to very high levels.
To see that this conjecture is a non trivial one, let us compute the character
 of the space of local fields eq.(\ref{loc1}). Attributing the degree 2 to
 $u$ and 1 to $\partial_1$, we find that
\begin{eqnarray}
\chi_1 &=& \prod_{j \geq 2} {1 \over 1 - p^j} = 
(1-p) \prod_{j \geq 1} {1 \over 1 - p^j}
= 1 +p^2 +p^3 +2p^4 +2p^5 +\cdots
\nonumber
\end{eqnarray}
On the other hand the character of the elements in the right hand side
 of eq.(\ref{loc2}) is
\begin{eqnarray}
\chi_2 &=& \prod_{j \geq 1} {1 \over 1 - p^{2j-1}}  \prod_{j \geq 1} {1 \over 1 - p^{2j}}
= \prod_{j \geq 1} {1 \over 1 - p^j}
= 1 +p +2p^2 +3p^3 +5 p^4 +7p^5 +\cdots
\nonumber
\end{eqnarray}
Hence $\chi_1 < \chi_2$, and this is precisely why null-vectors exist.
Let us give some examples of null-vectors
\begin{eqnarray}
level~1&:& \partial_{1}\cdot 1 =0 \label{nvcl} \\
level~2&:& \partial^2_{1}\cdot 1 =0 \nonumber \\
level~3&:& \partial^3_{1}\cdot 1 =0, \quad\partial_{3}\cdot 1 =0 \nonumber \\
level~4&:& \partial^4_{1}\cdot 1 =0,
\quad \partial_{1}\partial_{3}\cdot 1 =0,
\quad(\partial^2_{1} S_2 -4 S_4 +6 S_2^2)\cdot 1 =0 \nonumber \\
level~5&:& \partial^5_{1}\cdot 1 =0,\quad\partial_{1}^2\partial_{3}\cdot 1 =0,
\quad\partial_{5}\cdot 1 =0, \nonumber \\
&&\partial _{1}(\partial^2_{1} S_2 -4 S_4 +6 S_2^2)\cdot 1 =0,
\quad (\partial_{3} S_2 -\partial_{1} S_4 )\cdot 1 =0 \non
\end{eqnarray}
We have written all the null-vectors
explicitly to show that their numbers exactly
match the character formulae.
The non trivial null-vector at level 4 expresses $S_4$
in terms of the original variable $u$:
$4 S_4=-{1\over 2}u'' +{3\over 2} u^2 $. With this identification
the non-trivial null-vector at level 5, $\partial_{3} S_2 -\partial_{1} S_4$,
gives the KdV equation itself
\begin{eqnarray}
\partial_{3} u +{3\over 2}uu' -{1\over 4} u''' =0
\nonumber
\end{eqnarray}

More generally one can consider
the descendents of the fields
$e^{im \varphi}$ where $\varphi$ is related to $u$ by the Miura transformation
$$ u = -\varphi'^2 +i\varphi'' $$
Here, the presence of $i$ is a matter of convention. The reality problems
have been discussed at length in \cite{bbs}.

For this consideration and for other purposes we need certain information
about the Baker-Akhiezer function.
The Baker-Akhiezer function $w(t,A)$ is a
solution of the equation
\begin{eqnarray}
L w(t,A) = A^2 w(t,A)
\label{baker}
\end{eqnarray}
which admits an asymptotic
 expansion at $A = \infty$ of the form
\begin{eqnarray}
w(t,A) = e^{\zeta(t,A)}( 1 + 0(1/A));~~~ \zeta(t,A) = \sum_{k\geq 1} t_{2k-1} A^{2k-1}
\nonumber
\end{eqnarray}
In these formulae, higher times are considered as parameters.
The second solution of equation (\ref{baker}), denoted
by $w^*(t,A)$, has the asymptotics
\begin{eqnarray}
w^*(t,A) = e^{-\zeta(t,A)}( 1 + 0(1/A))
\nonumber
\end{eqnarray}
These definitions do not fix completely the Baker-Akhiezer
functions since we can
still multiply them by constant
asymptotic series of the form $1+O(1/A)$. Since
normalizations will be important to us,
let us give a more precise definition.
We first introduce the dressing operator $\Phi$:
\debut
L = \Phi \partial_1^2 \Phi^{-1};
\quad \Phi = 1 +\sum_{i>1} \Phi_i \partial_1^{-i}
\non
\fin
and we define
\debut
w(t,A) = \Phi e^{\zeta(t,A)},\quad w^*(t,A) = (\Phi^*)^{-1} e^{-\zeta(t,A)}
\non
\fin
where $\Phi^* = 1 +\sum_{i>1} (-\partial_1)^{-i} \Phi_i$ is the formal adjoint of
$\Phi$.
\proclaim Proposition 10.
With the above definitions, one has \hfill \break
\noindent
1) The wronskian $W(A)=w(t,A)' w^*(t,A) - w^*(t,A)'w(t,A)$ takes the value
\debut
W(A) = 2 A
\non
\fin
2) The generating function of the local densities $S(A)=1+\sum_{k>0}
S_{2k}A^{-2k}$ is related to the Baker-Akhiezer function by
\debut
S(A) = w(t,A) w^*(t,A)
\non
\fin
3) The function $S(A)$ satisfies the Ricatti equation
\debut
2S(A)S(A)''-(S(A)')^2-4uS(A)^2-4A^2S(A)^2+4A^2=0
\label{ric}
\fin
\par
\proof
Let us prove the wronskian identity. This amounts to showing that
$res_A (W(A) A^i)= 2 \delta_{i,-2}$. But we have
\debut
res_A \Big(W(A) A^i\Big) =
res_A \left\{ \Big(\partial _1\Phi \partial _1^i e^{\zeta(t,A)} \Big)
\Big( (\Phi^*)^{-1} e^{-\zeta(t,A)} \Big)
- \Big(\Phi e^{\zeta(t,A)}\Big)\Big(\partial _1 (\Phi^*)^{-1}
(-\partial _1)^i e^{-\zeta(t,A)}\Big)\right\}
\non
\fin
We can transform the residue in $A$ in a residue in
$\partial _1$ using the formula
\debut
res_A \left\{ \Big(P e^{\zeta(t,A)}\Big)\cdot \Big(Qe^{-\zeta(t,A)}\Big) \right\}
= res_{\partial _1}
\Big(PQ^* \Big)
\non
\fin
Hence we find
\debut
res_A \Big(W(A) A^i\Big) = res_{\partial _1}
\Big\{ \partial _1\Phi \partial _1^i \Phi^{-1}
+ \Phi \partial _1^i \Phi^{-1} \partial _1\Big\}
= res_{\partial_1} \Big\{ \partial_1 L^{i\over 2} +  L^{i\over 2} \partial_1
\Big\} \non \fin
If $i$ is even positive, the residue is zero because the $L^{i\over 2}$ is a
purely differential operator. If $i=-2$ the residue is obviously $2$, and if
$i<-2$ it is zero.  If $i$ is odd, then $(L^{i\over 2})^*=-L^{i\over 2}$ so that
the operator $\partial _1 L^{i\over 2} +  L^{i\over 2} \partial_1$ is formally
self-adjoint and it cannot have a residue.

The proof of 2) is simple \cite{dickey}
\debut
res_A \Big(A^{2k-1} w(t,A)w^*(t,A) \Big) =
res_{\partial _1}\Big(\Phi \partial _1^{2k-1}\Phi^{-1}\Big) =
res_{\partial _1}\Big(L^{2k-1\over 2}\Big) = S_{2k}
\non
\fin

The Ricatti equation follows immediately from 1),2) and eq.(\ref{baker}).
\square

Let us return to the descendents of the primary fields.
For the descendents of the fields $e^{im \varphi}$, our conjecture states
that
\begin{eqnarray}
\CO (u,u',u'',...)e^{im \varphi} &=&  \sum_{\nu \geq 0}\partial^\nu
\left( F_{\CO,\nu} (S_2, S_4, \cdots) e^{im \varphi} \right)
\label{loc3}
\end{eqnarray}

Let us consider for example $e^{ i\varphi}$.
For a true solution of the KdV equation, the Baker-Akhiezer function is a true
function on the spectral curve, and it can be analytically continued at $A =0$.
From the definition of $e^{i\varphi}$ we have $L e^{i\varphi} =0$.
Comparing with eq.(\ref{baker}), we see that $e^{i\varphi} = w(t,A)\vert_{A=0}
$.  To check eq.(\ref{loc3}),
at least on the first few levels, we need the time
derivatives of $e^{i\varphi}$. They can be obtained as follows.
The time evolution of the Baker-Akhiezer function is well known.
\begin{eqnarray}
{\partial w \over \partial t_{2k-1} } = \left( L^{2k-1\over 2} \right)_+ w
\nonumber
\end{eqnarray}
By analytical continuation at $A=0$,
we obtain the evolution equations for $e^{i\varphi}$.

Let us give some examples of these null vectors. We show below the
first null vector associated to the primary fields $e^{im \varphi}$.
\begin{eqnarray}
m=1&:& (\partial_{1}^2+2 S_2)e^{i\varphi} =0 \non \\
m=2&:& (2 \partial_3 + \partial_1^3 + 6 \partial_1 S_2)e^{2i\varphi} =0
\nonumber \\ m=3&:& (8 \partial_1 \partial_3 + \partial_1^4 + 12 \partial_1^2
S_2 + 24 S_4)e^{3i\varphi} =0
\nonumber \\ m=4&:& ( 24 \partial_5 \partial_1 +20
\partial_3 \partial_1^3+ \partial_1^6 + 20 \partial_1^4 S_2 + 40 \partial_3
\partial_1 S_2 + 120 \partial_1^2 S_4) e^{4i\varphi}=0 \non
\end{eqnarray} In these formulae, the derivatives act on everything on their
right i.e.  $ \partial_1 S_2 e^{2i\varphi} =  \partial_1 ( S_2 e^{2i\varphi})$.

\subsection{Finite-zone and soliton solutions.}

For the finite-zone solutions, the Baker-Akhiezer function is
an analytical function on the spectral curve which is an
algebraic Riemann surface.
Let us recall briefly the construction \cite{itsmat,novbook}.

We start with an hyperelliptic curve
$\Gamma$ of genus $n$ described by the equation
\debut
\Gamma : Y^2 &=& X {\cal P}(X),~~ {\cal P}(X) = \prod_{j=1}^{2n}(X
-B_j^2) ,\quad B_{2n}>\cdots >B_2>B_1>0\non
\fin
For historical reasons we prefer to work with the parameter $A$
such that $X=A^2$.
The surface is realized as the $A$-plane with cuts
on the real axis over the intervals
$c_i=(B_{2i-1}, B_{2i})$ and $\bar{c}_i=(-B_{2i}, -B_{2i-1})$,
$i=1,\cdots, n$,
the upper (lower) bank of $c_i$ is identified with
the upper (lower) bank of $\bar{c}_i$.
The  square root $\sqrt{\CP(A^2)}$ is chosen so that
$\sqrt{\CP(A^2)}\to A^{2n}$ as $A\to\infty$.
The canonical basis of cycles is chosen as follows:
the cycle $a_i$ starts from $B_{2i-1}$ and goes
in the upper half-plane
to $-B_{2i-1}$, $b_i$ is an anti-clockwise cycle around the cut $c_i$.

Let us consider in addition a divisor of order $n$ on the surface $\Gamma$:
$$\CD =(P_1,\cdots ,P_n)$$
With these data we construct the Baker-Akhiezer function which is the unique
function with the following analytical properties:
\begin{itemize}
\item It has an essential singularity at infinity:
$w(t,A) = e^{\zeta(t,A)}(1 + O(1/A))$.
\item It has $n$ simple poles outside infinity. The divisor of these
poles is $\CD$.
\end{itemize}
Considering the quantity $-\partial_1^2 w + A^2 w$, we see that
it has the same analytical properties as $w$ itself, apart for the first
normalization condition. Hence, because $w$ is unique,
 there exists a function $u(t)$ such that
\debut
-\partial_1^2w + u(t) w + A^2 w =0
\label{A0}
\fin
We recognize eq.(\ref{baker}).
One can give various explicit constructions of the Baker-Akhiezer
 function.
Let us introduce the divisor  ${\cal Z}(t)$
of the zeroes of the Baker-Akhiezer
function. It is of degree $n$:
\debut
\CZ(t)  = (A_1 (t), \cdots, A_n(t) ) \non
\fin
The equations of motion for the divisor $\CZ(t)$ read \cite{novbook}.
\debut
\partial_1
A_i(t) = -{  \sqrt{ {\cal P}(A_i^2(t))} \over \prod\limits_{j \neq i}
(A_i^2(t) - A_j^2(t) ) }
\label{A5}
\fin
The normalization of the Baker-Akhiezer function corresponds to
a particular choice of the divisor of its poles $\CD$.
Later we shall specify the divisor which corresponds to the
normalization of the Baker-Akhiezer function which
was required in the previous subsection, for the moment we
give a formula in which the normalization is irrelevant.
Consider two sets of times $t$ and $t^{(0)}$, differing only by
the value of $t_1$. Then we can write
\debut
{w(t,A)\over w(t^{(0)},A)} =
\sqrt{ Q(A^2,t) \over Q(A^2, t^{(0)}) }
\exp \left( { \int_{t_1^{(0)}}^{t_1}
 { A\sqrt{{\cal P}(A^2)} \over Q(A^2, t)} dt_1 }\right)
\label{A6}
\fin
where the polynomial $Q(A^2,t)$ is defined as $Q(A^2,t) = \prod_i
(A^2 -A_i^2(t))$.

The ratio of two dual Baker-Akhiezer functions ${w^*(t,A)\over
w^*(t^{(0)},A)}$ is obtained by
applying the hyperelliptic involution. This amounts to the
reflection $A\to -A$ in eq.(\ref{A6}).
Let us prove the following simple proposition.

\proclaim Proposition 11.
For the Baker-Akhiezer functions $w(t,A)$, $ w^*(t,A)$ normalized
by
$$w(t,A)' w^*(t,A) - w^*(t,A)' w(t,A)=2A $$
we have
\begin{eqnarray}
S(A) = {Q(A^2) \over \sqrt{{\cal P}(A^2)} }\equiv
 \exp \left(- \sum_k {1\over k} J_{2k}A^{-2k} \right)
\label{deffJ}
\fin
the latter equality is the definition of $J_{2k}$.
We recall that $Q(A^2)$ and ${\cal P}(A^2)$ are the polynomials
\begin{eqnarray}
Q(A^2) =\prod\limits_{i=1}^n (A^2 - A_i^2),~~
{\cal P}(A^2) = \prod\limits_{i=1}^{2n} (A^2 - B_i^2) \nonumber
\end{eqnarray}
\par
\proof
To prove the proposition we use the Wronskian identity
\debut
w(t,A)' w^*(t,A) - w^*(t,A)'  w(t,A)
= w(t,A) w^*(t,A)
\partial_1\( \log {w(t,A) \over w^*(t,A)}\) =2A
\non
\fin
but using eq.(\ref{A6}) and the fact
that ${w(t,A)\over w(t^{(0)},A)}$ and ${w^*(t,A)\over w^*(t^{(0)},A)}$
differ by the
sign of the square root, we have
\debut
\partial_1\( \log {w(t,A) \over w^*(t,A)} \)
= 2 {A \sqrt{{\cal P}(A^2)} \over Q(A^2)}
\non
\fin
and the result follows.\square

Notice that for $J_{2k}$ defined in (\ref{deffJ}) we have
\begin{eqnarray}
J_{2k} = \sum_i A_i^{2k} -{1\over 2} \sum_i B_i^{2k}
\nonumber
\end{eqnarray}

   From eq.(\ref{deffJ}) we see that the normalization of
$w(t,A)$ and $w^*(t,A)$ which corresponds to the proper value
of the Wronskian is such that the divisors $\CD$ and $\CD ^*$
are composed of Weierstrass points and  $\CD +\CD ^*=
(B_1,\cdots ,B_{2n})$. Actually,
it is this quite unique normalization which was used by Akhiezer
in his original paper.

Now we are in position to describe the dynamics of $S(A)$ with
respect to all times.
It is very useful to define the following strange object
\begin{eqnarray}
dI(D) = \sum_{k \geq 1} D^{-2k}{\partial \over \partial t_{2k-1}}dD
\label{defI1}
\end{eqnarray}
$dI(D)$ is a 1-form in the  $D$-plane and a vector field with respect to times.
We have
\proclaim Proposition 12.
\begin{eqnarray}
dI(D)\cdot S(A) =   { S(D) S(A)' - S(A) S(D)'
 \over D^2 - A^2 }dD
\label{IS}
\end{eqnarray}
\par
\proof
We give a proof of this proposition for the finite zone solutions, which
are our main concern here, but clearly the formula is quite
general. We are sure that a general proof of equation (\ref{IS})
exists, but it must be based on manipulations with asymptotic
formulae. We prefer to work with analytical functions.
Anyway, every solution of KdV can be obtained from the finite-zone ones
by a suitable limiting procedure, so considering finite-zone
solutions is not a real restriction.

Let us describe the motion, under the time $t_{l}$, of the divisor
$\CZ(t) $ of the zeroes of the Baker-Akhiezer function.
Introduce the normalized holomorphic differentials $d\omega _i$ for
$i=1,\cdots ,n$
and the normalized second kind differentials with singularity at
infinity $d\tilde{\omega} _{2i-1} $, $i\ge 1$
\debut
&&\int\limits _{a_j}d\omega _i =\delta _{i,j}\non\\
&&\int\limits _{a_j} d\tilde{\omega} _{2i-1}=0,\qquad
d\tilde{\omega} _{2i-1}(A)=d(A^{2i-1})+\CO(A^{-2})dA\ for \ A\sim\infty
\non
\fin
It is well known that, by the Abel map, this motion is transformed into a
linear flow on the Jacobi variety.
\debut
{\partial \over \partial t_{2l-1}} \sum_j \int^{A_j} d\omega_k
= \int_{b_k} d\tilde{\omega}_{2l-1}
\non
\fin
Form this equation one easily find~:
\debut
dI(D)\cdot \sum_j \int^{A_j} d\omega_k
= \int_{b_k} d\tilde{\omega}_{D}
\label{aaa}
\fin
where $d\tilde{\omega}_{D}$ is a 2-differential
defined on $\Gamma \times \Gamma _0$
($\Gamma _0$ is the Riemann sphere)
parametrized by $A$ and $D$ respectively.
It is useful to think of $\Gamma _0$ as a realization
of the curve $Y^2=X$ similar to $\Gamma$.
The  $a$-periods  of $d\tilde{\omega}_{D}$ on $\Gamma$ vanish.
The only singularities of the differential $d\tilde{\omega}_{D}$ are
the second order poles
at the two points $A=\pm D$:
$$d\tilde{\omega}_{D} (A)=\Bigl({A^2+D^2\over (A^2-D^2)^2}+\CO (1)\Bigr)dA\,dD$$
By Riemann's bilinear relations one easily find:
$$\int_{b_k} d\tilde{\omega}_{D}=d\omega_k(D)$$
Eq.(\ref{aaa}) then takes the form
\debut
dI(D)\cdot \sum_j \int^{A_j} d\omega_k
= d\omega_k(D)
\label{aaaa}
\fin
The normalized holomorphic differentials are linear combinations of the
differentials
\debut
d\sigma_k(A) = {A^{2k-2} \over \sqrt{{\cal P}(A^2)}}dA,~~k=1\cdots n
\non
\fin
with coefficients depending on $B_i$. They do
not depend on times. Hence by linearity
we can write for $d\sigma_k $ the same equation as (\ref{aaaa}).
Differentiating explicitly we get
the following system of equations
\debut
\sum_{j =1}^n {A_j^{2k-2} \over  \sqrt{{\cal P}(A_j^2)}}dI(D)\cdot A_j =
 {D^{2k-2} \over
\sqrt{{\cal P}(D^2)}}dD; \qquad k=1\cdots n
\non
\fin
Solving this linear system of equations gives
\debut
dI(D)\cdot A_j^2 =  2A_j~ {Q(D^2) \over  \sqrt{{\cal P}(D^2)}}
{1\over D^2-A_j^2}{\sqrt{\CP(A_j^2)}\over
\prod_{i \neq j}(A_j^2-A_i^2)} dD= - {Q(D^2) \over  \sqrt{{\cal P}(D^2)}}
{1\over D^2-A_j^2}\partial _1 A_j^2\ dD
\non
\fin
where we have used eq.(\ref{A5}) in the last step.
Finally, using eq.(\ref{deffJ}) we find
\debut
dI(D)\cdot S(A) =  S(D)S(A)\sum_j {1\over D^2 -A_j^2}
{1\over A^2 - A_j^2}
\partial_1(A_j^2) dD
= {1\over D^2 - A^2} S(D)S(A)\left( \log {S(D)\over S(A)}\right)'~dD
\non
\fin
\square

The soliton solutions correspond to
a rational degeneration of the
finite-zone solutions such that
$$B_{2j-1}\to \tilde{B}_j \leftarrow B_{2j}, \qquad j=1,\cdots ,n$$
The points of the divisor $A_i$ and the points $\tilde{B}_j$
are the coordinates in the $n$-solitons phase space.
In \cite{bbs} we gave a detailed discussion of the Hamiltonian
structure. In particular we have
\debut
{\partial \over \partial t_{2k-1}} =
\{ I_{2k-1},~~\cdot \}; \quad I_{2k-1}= {1\over 2k-1}\sum_i \tilde{B}^{2k-1}_i
\non
\fin
The expressions for the local observables also
 follow easily from the finite-zone case. In particular
\debut
J_{2k}(A,\tilde{B})=\sum\limits _{i=1}^n\Bigl( A_i^{2k}-\tilde{B}_i^{2k}\Bigr)
\label{solJ}
\fin
The expressions for $S_{2k}$ follow from here.

\subsection{ Classical limit of $\CQ$ and $\CC$.}

Let us consider the classical limit $\nu\to\infty$
 of the operators $\CQ$ and $\CC$.
To this aim, we have to understand the relation between the quantum
and the classical descriptions of the observables. In the quantum case we
considered the form factors i.e. matrix elements of the form
$$f_{\CO}(\b _1, \cdots ,\b _{2n})_{-\cdots -+\cdots +}=
\langle 0|\CO (0)|\b_1,\cdots ,\b _n;\ \b_{n+1},\cdots ,\b _{2n}\rangle$$
where $\b_1,\cdots ,\b _n$ are rapidities of anti-solitons,
$\b_{n+1},\cdots ,\b _{2n} $ are rapidities of solitons.
The matrix elements of this form do not allow a direct
semi-classical interpretation, it is necessary to perform a crossing
transformation to the matrix elements between two $n$-soliton states:
\debut
&&\langle \b_1,\cdots ,\b _n|\CO (0)|\b_{n+1},\cdots ,\b _{2n}\rangle =
f_{\CO}(\b _1-\pi i, \cdots ,\b _{n}-\pi i,
\b _{n+1}, \cdots ,\b _{2n})_{-\cdots -+\cdots +}
\non\fin
In \cite{bbs} it is explained that the formula (\ref{ints})  for
this form factor is a result of quantization of $n$-soliton solutions
in which $A_1,\cdots ,A_n$ play the role of coordinates,
$B_1,\cdots ,B_n$ and $B_{n+1},\cdots ,B_{2n}$ give the
collection of eigenvalues for two eigenstates.

Recall that the generating function for the local descendents
of the primary field $\Phi _m$ was
written as follows
\debut
\CL _m(t,y|A|B)=
\exp\Bigl(\sum\limits _{k\ge 1}t_{2k-1}I_{2k-1}(B)
+y_{2k}J_{2k}(A|B)\Bigr)~~
\({\prod\limits _{i=1}^n A_i^m
\prod\limits _{j=1}^{2n} B_j^{-{m\over 2}} }\) \label{qqg}
\fin
where $I_{2k-1}(B)$ and $J_{2k}(A|B)$ are defined in eqs.(\ref{defI},\ref{defJ}).
The expression
$$
\exp\Bigl(\sum\limits _{k\ge 1}y_{2k}J_{2k}(A|B)\Bigr)
\({\prod\limits _{i=1}^n A_i^m
\prod\limits _{j=1}^{2n} B_j^{-{m\over 2}} }\) \non
$$
is practically unchanged under the crossing transformation which
corresponds to $B_i\to -B_i$ for $i=1,\cdots ,n$. Comparing it with
the classical formulae (\ref{solJ}) we see that it corresponds to
special symmetric ordering of them, for example
$$J_{2k}(A|B)={1\over 2}\Bigl(
J_{2k}(A_1,\cdots ,A_n,B_1,\cdots ,B_n )+
J_{2k}(A_1,\cdots ,A_n,B_{n+1},\cdots ,B_{2n})\Bigr)$$
This ordering is a prescription which we make for the quantization.

On the other hand the eigenvalues of the Hamiltanians $I _{2k-1}(B)$ under
crossing transformation change to
$$-I _{2k-1}(B_1,\cdots ,B_n) +I _{2k-1} (B_{n+1},\cdots ,B_{2n}) $$
i.e. the descendents with respect to $I _{2k-1}$ correspond to
taking commutator of $\CO$ with $I _{2k-1}$. Certainly the classical
limit makes sense only for the states with close eigenvalues,
so, it is needed that
$$s_{2k-1}(B_1,\cdots ,B_n)-s_{2k-1}(B_{n+1},\cdots ,B_{2n})=\CO(\xi)$$
Recall that $\xi =-i\log (q)$ plays the role of Plank's constant.

Thus comparing the classical and quantum pictures provides the following
result. The quantum generating function (\ref{qqg}) corresponds to the classical
generating function
\debut
\CL _m^{cl}(t,y)=
\exp\Bigl(\sum\limits _{k\ge 1}t_{2k-1}I_{2k-1}\Bigr)\cdot
\exp\Bigl(\sum\limits _{k\ge 1}y_{2k}J_{2k}\Bigr)\ e^{im\varphi} \label{clg}
\fin
where $\cdot$ means the application of Poisson brackets.
In fact $I_{2k-1}$ can be replaced by $\partial _{2k-1}$.
The normalization in the formula for $I(B)$ (\ref{defI})
is chosen in order to provide an exact agreement with the classical formulae.

Now let us consider the classical limit of the operators $\CQ$ and
$\CC$
in the Neveu-Schwarz sector. For $\CQ$ we had the formula
\debut
\CQ=
\int\limits {dD\over D}
e^{X(D)} \psi(D)= \int\limits {dD\over D} \sinh(X(D)) \psi(D)
\non
\fin
The latter equation is due to the fact that the fermion is odd.
From the definition (\ref{defXD}) of $X(D)$, we have
in the classical limit
$$X(D)\to\  \frac{-i\xi}{2}
\sum\limits _{k\ge 1}D^{-2k+1}I_{2k-1}$$
Hence the following expression is finite in the classical limit:
\debut
\CQ _{cl}= \ \lim_{\xi\to 0} \ i2{\CQ\over \xi}=\int \psi (D)dI(D)
\label{qcl}
\fin
where $dI(D)$ is the 1-form in $D$-plane introduced in the previous
subsection:
$dI(D) =\sum\limits _{k\ge 1}D^{-2k}I_{2k-1} dD$. Remark that $\CQ_{cl}$
can be thought of as a generalized Dirac operator.

For $\CC$ we had the formula
\debut
&&\CC=
\int\limits _{|D_2|>|D_1|}{dD_2\over D_2} \int\limits {dD_1\over D_1}
e^{X(D_1)}e^{X(D_2)}\tau _e \left({D_1\over D_2}\right)
\psi(D_1)\psi(D_2) =\non\\ &&=
\int\limits _{|D_2|>|D_1|}{dD_2\over D_2} \int\limits {dD_1\over D_1}
\cosh (X(D_1))\cosh (X(D_2))\tau ^- _e \left({D_1\over D_2}\right)
\psi(D_1)\psi(D_2)+\non\\&&+
\int\limits _{|D_2|>|D_1|}{dD_2\over D_2} \int\limits {dD_1\over D_1}
\sinh (X(D_1))\sinh(X(D_2))\tau ^+_e \left({D_1\over D_2}\right)
\psi(D_1)\psi(D_2) \non
\fin
where $ \tau_e ^+ $ and $\tau ^-_e $ are even and odd parts of $\tau _e$:
$$\tau ^-_e(x)=\sum\limits _{k=1}^{\infty}{1-q^{2k-1}\over 1+q^{2k-1}}
\ x^{2k-1},
\qquad
\tau ^+_e(x)=-\sum\limits _{k=1}^{\infty}{1+q^{2k}\over 1-q^{2k}}\ x^{2k} $$
Obviously when $\xi\to 0$ one has
$$  \tau ^-_e(x)\to \frac{-i\xi}{2}\ x{d\over dx}\({x\over 1-x^2}\),
\quad  \tau ^+_e(x)\to -(i\xi)^{-1}\log (1-x^2)$$
So, the following expression is finite in the classical limit
\debut
&&\CC _{cl}= \ \lim_{\xi \to 0} \ {2\CC\over \pi\xi} =
\non\\
&&=
\int \psi(D){d\over dD}\psi(D)dD +{1\over 2\pi i}
\int\limits _{|D_2|>|D_1|}dI(D_2) \int dI(D_1)
\log\left(1-\({D_1\over D_2}\)^2\right)
\psi(D_1)\psi(D_2)\quad
\label{ccl}
\fin
In the next subsection we are going to apply these operators to
description of the classical KdV hierarchy. Notice that
as usual the quantum formulae
are far more symmetric than the classical ones.

\subsection{The classical equations of motion from $\CQ_{cl}$ and $\CC _{cl}$.}

In this subsection we shall consider only the descendents
of the identity, i.e. the pure KdV fields.
We have described this  space by the generating function (\ref{clg}):
\debut
\CL _{m=0}^{cl}(t,y)=
\exp\Bigl(\sum\limits _{k\ge 1}t_{2k-1}I_{2k-1}\Bigr)\cdot
\exp\Bigl(\sum\limits _{k\ge 1}y_{2k}J_{2k}\Bigr)\cdot~1 ,
 \non
\fin
Let us fermionize $J_{2k}$ and apply the equations
\debut
\langle \Psi _{-3}|\ \CQ_{cl}\w= 0,\qquad
\langle \Psi _{-5}|\ \CC _{cl}\w= 0
\label{cqcleq}
\fin
to the description of the equations of motion.
In this section the symbol $\w= 0$ means the vanishing of the
scalar product with the generating of local fields.

We give the list
of null-vectors following from these two equations
up to the level 5 specifying explicitly the vectors $\langle \Psi _{-3}| $
and $\langle \Psi _{-5}|$ from which they come. We do not write the
descendents with respect to $I$'s of the already listed null vectors:
\newline Null vectors coming from $\CQ_{cl}$:
\debut
\langle -1 \vert \psi_{-1}^* &:& \partial_1 \cdot 1 \non \\
\langle -1 \vert \psi_{-3}^* &:& \partial_3 \cdot 1 \non \\
\langle -1 \vert \psi_{-5}^* &:& \partial_5 \cdot 1 \non \\
\langle -1 \vert \psi_{-3}^*\psi^*_{-1}\psi_1 &:& (-\partial_3 S_2 +
\partial_1 S_4) \cdot 1 \non
\fin
\newline Null vectors coming from $\CC _{cl}$:
\debut
\langle -1 \vert \psi_{-3}^*\psi^*_{-1} &:& (\partial_1^2 S_2 - 4 S_4 + 6
S_2^2 + {1\over 2} \partial_1 \partial_3 ) \cdot 1 \non
\fin

Obviously these null-vectors coincide with (\ref{nvcl}).
So, in particular, equations  (\ref{cqcleq}) imply the KdV
equation itself. We have verified that the null-vectors coincide with
those obtained from the Gelfand-Dickey construction up to
level 16. On higher levels we find higher equations
of KdV hierarchy.

We have seen that the KdV equation follows from the equations
(\ref{cqcleq}). Let us prove the opposite: equations
(\ref{cqcleq}) hold on any solution of KdV.
We start with the operator $\CQ_{cl}$.
\proclaim Proposition 13.
Let
\debut
{\CQ _{cl}}&=& \int \psi(D) dI(D) = \sum_{k \geq 1}
\psi_{-2k+1}{\partial \over \partial t_{2k-1}}
\non
\fin
Then if $J_{2k}$ are constructed from a solution of KdV we have
\debut
{\cal Q }_{cl}\exp \left( -\sum_{k \geq 1} {1\over k} J_{2k} h_{-2k} \right)
\vert -1 \rangle &=& 0
\non
\fin
\par
\proof
Let us introduce the notation:
$$T=\exp \left( -\sum_{k \geq 1} {1\over k} J_{2k} h_{-2k} \right)
= \exp \left( \int {dA\over 2i\pi A} \log S(A) h(A)\right) $$
where
$h(A)=\sum\limits _{k\ge 1}h_{-2k}A^{2k}$.
We have $ \psi(D)\ T  = T \ S^{-1}(D) \psi(D) $. So
\debut
{\cal Q }_{cl}\ T\  \vert -1 \rangle
&=& T
\int {1\over S(D) S(A)} \psi(D) h(A)
{dA \over 2i\pi A}(dI(D)\cdot S(A))
\vert -1 \rangle
\non
\fin
We now use eq.(\ref{IS}) to get
\debut
&&\int {1\over S(D) S(A)} \psi(D) h(A)
{dA \over 2i\pi A}(dI(D)\cdot S(A))\vert -1 \rangle =\non\\
&&\hskip 1cm
= \int dD  {dA \over 2i\pi A}{1 \over D^2 - A^2} \left[
\(\log S(A)\)' - \(\log S(D)\)' \right]
 \psi(D) h(A) \vert -1 \rangle
\non
\fin
But
\debut
 \psi(D) h(A) \vert -1 \rangle = : \psi(D)\psi(A)\psi^*(A): \vert -1 \rangle
+ {A D\over D^2 - A^2} \psi(A) \vert -1 \rangle ~~~ |D| > |A|
\non
\fin
Let us consider first the integral
\debut
\int_{|D|>|A|} {dD \over D} {dA \over 2i\pi A}{D \over D^2 - A^2}
\(\log S(A)\)'
\left[: \psi(D)\psi(A)\psi^*(A): - {A D\over D^2 - A^2} \psi(A)
 \right]\vert -1 \rangle
\non
\fin
One can do the integral over $D$. Notice that the integrand is regular
at $D=0$. Hence,  the contributions to the integral come from the poles at
 $D^2 = A^2$. The simple pole does not contribute because its residue vanishes
since we have the product of  two fermion fields at the same point in the normal
product. The  double pole obviously does not contribute either, and the integral
is zero. Next we look at the integral
\debut
\int_{|D|>|A|} {dD \over D} {dA \over 2i\pi A}{D \over D^2 - A^2}
\(\log S(D)\)'
\left[: \psi(D)\psi(A)\psi^*(A): - {A D\over D^2 - A^2} \psi(A)
 \right]\vert -1 \rangle
\non
\fin
This time one can do the integral over $A$. But it is clear that the integrand
is regular at $A=0$, and the integral also vanishes. \square

Let us consider now the operator $\CC _{cl}$.
\proclaim Proposition 14.
Let
\debut
{\cal C }_{cl}&=& \int dD \psi(D) {d \over dD} \psi(D)
+{1\over 2i\pi} \int_{|D_1|<|D_2|}
\log\left(1 -{D_1^2 \over D_2^2}\right) \psi(D_1) \psi(D_2)d I(D_1)d I(D_2)
\non
\fin
then if $J_{2k}$ are constructed from a solution to KdV
\debut
{\cal C }_{cl}\exp \left(- \sum_{k \geq 1} {1\over k} J_{2k} h_{-2k} \right)
\vert -1 \rangle = 0 \non
\fin
\par
\proof
Let us split ${\cal C }_{cl}$ into two pieces ${\cal C }_{cl}={\cal C}_1 +
{\cal C}_2$:
\debut {\cal C}_1 &=&  \int {dD\over D} \psi(D) D {d \over dD} \psi(D)
\non \\ {\cal C}_2 &=&
{1\over 2\pi i}\int_{|D_1|<|D_2|}
\log\left(1 -{D_1^2 \over D_2^2}\right)  \psi(D_1)  \psi(D_2)dI(D_1) dI(D_2)
\non
\fin
We use the same notation $T$ as in the previous proposition.
We treat first the ${\cal C}_2$ term
\debut
{\cal C}_2 \ T\ \vert -1 \rangle & =&- {1\over 2\pi i}
\int d I(D_2)\psi(D_2)\ T
\ \int\limits_{|A_1|<|D_1|<|D_2|}{d D_1 \over D_1}{d A_1 \over 2i\pi A_1}
\log \left(1 -{D_1^2 \over D_2^2}\right)
{D_1 \over D_1^2 - A_1^2 }\( \log {S(A_1)\over S(D_1)}\)^{\prime}
 \non \\
&&\hskip 4cm
\times\left[: \psi(D_1)\psi(A_1)\psi^*(A_1): - {A_1 D_1\over D_1^2 - A_1^2}
 \psi(A_1) \right]\vert -1 \rangle \non
\fin by the same argument as in the previous proof, we see that only the term
$\( \log S(A_1)\)'$ contributes. This time however, the double pole gives a
non vanishing contribution
 \debut
{1\over 2\pi i}\int\limits_{|D_1| > |A_1|}dD_1
\log\left(1 -{D_1^2 \over D_2^2}\right) {D_1 \over
(D_1^2 - A_1^2)^2 } = -{1\over D_2^2 - A_1^2} \non
\fin
so that
\debut {\cal C}_2 \ T\ \vert -1 \rangle & =&
- \int dI(D_2)\psi(D_2) \ T\ \int\limits_{|D_2| > |A_1|}
\frac{dA_1}{2i\pi} {\(\log S(A_1)\)' \over D_2^2 - A_1^2}
\psi(A_1) \vert -1 \rangle \non
\fin
Recall that in this formula $dI(D_2)$ acts on $T$ and on $(\log S(A_1))'$.
Commuting again $\psi(D_2)$ and $T$ we finally get
after having computed the integral over $D_2$
\debut {\cal C}_2\ T\ \vert -1 \rangle & =&
T\  \int {d A \over A^2}\left[
{1\over 2}( \log S(A))'' +{1\over 4} (( \log S(A))')^2 \right] \psi(A) {d \over
d A}\psi(A)\vert -1 \rangle \non
\fin
The calculation of the ${\cal C}_1$ term is straightforward
\debut {\cal C}_1 \ T\ \vert -1 \rangle & =&
\frac{T}{2}\  \int d A ~ S^{-2}(A)~  \psi(A)
{d \over d A}\psi(A)\vert -1 \rangle
\non \fin
Combining the two terms and using the Ricatti equation (\ref{ric})
\debut
2  (\log S(A))'' + (( \log S(A))')^2 +4 A^2 S^{-2}(A) = 4
(u +A^2)
\non
\fin
we get
$$ \CC _{cl} \ T\ \vert -1 \rangle =\int {dA\over A^2}(u+A^2)
\psi (A){d \over dA}\psi(A)\vert -1 \rangle =0$$
the latter integral vanishes since the integrand is regular. \square

The equation (\ref{cqcleq}) for the generating function $\CL^{cl}(t,y)$,
defined by
\debut
\CL^{cl}(t,y)= \exp\({-\sum_{k \geq 1} y_{2k} J_{2k}(t) }\) \non
\fin
can be written explicitely by bosonizing the fermions. The
equation $\bra{\Psi_{-3}} \CQ_{cl}=0$ then reads:
\debut
\int D~ e^{\xi(D,y)}~dI(D) \CL^{cl}(t,y-[D]_e) = 0 \label{newQcl}
\fin
where $dI(D)= \sum_k D^{-2k}\d_{2k-1}dD$ as before
(it acts on $\CL ^{cl}$ by $\partial _{2k-1}$) and,
\debut
\xi(D,y)=\sum_{k\geq 1} D^{2k} y_{2k} \quad;\quad
[D]_e= ( D^{-2},\cdots, \frac{D^{-2k}}{k},\cdots) \non
\fin
The  $\bra{\Psi_{-5}} \CC_{cl}=0$ equation reads:
\debut
&&8\pi i\int dD D^3 ~ e^{2\xi(D,y)}~ \CL^{cl}(t,y-2[D]_e)+ \label{newCcl}\cr
&&+ \int\limits _{|D_2|>|D_1|} D_2D_1 (D_1^2-D_2^2)
\log\({1- \frac{D_1^2}{D_2^2}}\) ~e^{\xi(D_1,y)+\xi(D_2,y)}
dI(D_1)dI(D_2) \CL^{cl}(t,y-[D_1]_e-[D_2]_e) = 0\non\cr
\fin
Equations (\ref{newQcl},\ref{newCcl}) provide a system of linear differential
equations for the Taylor coefficients of $\CL^{cl}(t,y)$. It
becomes a system of non-linear differential equations for the
$J_{2k}$ when the closure condition 
$\CL^{cl}(t,y)= \exp\({\sum_{k \geq 1} y_{2k} J_{2k}(t) }\)$ is
imposed.

Thus we have shown that the equations (\ref{cqcleq}) provide
the complete description of the KdV hierarchy. We find it
quite amazing. It is also interesting that this
new description came from
pure  quantum considerations.

\section{Connection with the Whitham method.}

There is an surprising relation between the methods of this
paper and the Whitham
equations for KdV \cite{whi,fla,kri,dn}. The present section is devoted to
the description of this relation.

Let us remind briefly what is the Whitham method about.
Suppose we consider the solutions of
KdV which are close to a given quasi-periodic solution.
The latter is defined by the set of ends of zones
$B_1^2,\cdots ,B_{2n}^2$. We know that for the finite-zone
solution the dynamics is linearized by the Abel transformation
to the Jacobi variety of the hyper-elliptic surface
$Y^2=X\CP (X)$ for $\CP (X)=\prod (X-B_j^2)$.
The idea of the Whitham method is to average over the fast
motion over the Jacobi variety and to introduce "slow times"
$T_j$ which are related to the original KdV times
as $ T_j=\epsilon t_j$ ($\epsilon \ll 1$), assuming that the
ends of zones $B_j$ become functions of these "slow times"
(recall that the ends of zones were the integrals of motion for the pure
finite-zone solutions).

For the given finite-zone solution the observables can be written in
terms of $\theta$-functions on the Jacobi variety, but this kind
of formulae is inefficient for writing the averages. One has
to undo the Abel transformation, and to write the observables in
terms of the divisor $\CZ=(A_1,\cdots ,A_n)$. The formulae for the
observables are much more simple in these variables, and the averages
can be written as abelian integrals, the Jacobian due to the Abel transformation
is easy to calculate. The result of this calculation is as follows \cite{fla}.
Every observable $\CO$ can be written as an even symmetric function
$L_{\CO}(A_1,\cdots ,A_n)$
(depending on $B$'s as on parameters), and for the average we have
$$\vev{\langle \ \CO\ \rangle}=\Delta ^{-1}
\int\limits _{a_1}{dA_1\over\sqrt{\CP(A^2_1)}}\cdots
\int\limits _{a_n}{dA_n\over\sqrt{\CP(A^2_n)}} L_{\CO}(A_1,\cdots ,A_n)
\prod\limits _{i<j}(A_i^2-A_j^2)$$
where
$$\Delta=det \(\int\limits _{a_i}{A^{2j-2}dA\over\sqrt{\CP(A^2)}}\)_
{i,j=1,\cdots n}$$
The similarity of this formula with the formula for the form factors
(\ref{ff}) is the first intriguing fact.
Indeed, we have the following dictionary. For the local observables, we
have
\debut
L_{\CO} \Longleftrightarrow L_{\CO}
\non
\fin
For the weight of integration, we have
\debut
{1\over \sqrt{{\cal P}(A)}} \Longleftrightarrow \prod_{j=1}^{2n} \psi(A,B_j)
\non
\fin
But the most striking feature is that the cycles of integration are replaced
by functions of $a_i = A_i^{2\nu}$
\debut
cycle~a_i \Longleftrightarrow a_i^{-i}
\non
\fin
The coincidence between the notations for $a_i$-variables and
$a_i$-cycles is therefore not fortuitous.
The explanation of the fact that the cycles are replaced by these functions is
given in \cite{bbs}, where it was shown that the factor 
$\prod_i a_i^{-i}$ selects the
classical trajectory in the semi-classical approximation of eq.(\ref{ints}).
So, the solution of a non trivial, full fledged, quantum field theory has
provided us with a very subtle definition of a {\it quantum} Riemann surface.

The main result of the Whitham theory is that
the averaged equations of motion can be written in the following form:
$$
{\partial \over \partial{T_{2p+1}}}d\tilde{\omega} _{2q+1}(A)=
{\partial \over \partial{T_{2q+1}}}d\tilde{\omega} _{2p+1}(A)
$$
where, as earlier,
$ d\tilde{\omega} _{2q+1},d\tilde{\omega} _{2p+1}$ are normalized
second kind differentials
with prescribed singularity at infinity:
$$ d\tilde{\omega} _{2l+1}(A)=d(A^{2l+1})+\CO(A^{-2})dA,
\qquad \int\limits _{a_j}
 d\tilde{\omega} _{2l+1} =0  $$
In fact every equation of this type contains many
partial differential
equations for $B_j$ because one can decompose it
with respect to the parameter $A$. To our knowledge these
equations have never been deduced directly from the averaging
integrals except for the special case $p=1, q=2$ which was
considered in the pioneering paper \cite{fla}. We want to show
that the methods of this paper allow to do that, also
we shall also explain how
to describe other equations of the hierarchy.

As it has been shown in this paper the classical observables can
be obtained from the generating function:
\debut
\exp (\sum_{k\geq 1} t_{2k-1}I_{2k-1})\cdot\exp (\sum_{k\geq 1} y_{2k}J_{2k})
\label{ave}
\fin
In the Whitham case the KdV times become fast which means that
$I_{2p-1}=\epsilon \widehat{I}_{2p-1}$ where $\widehat{I}_{2p-1}$
denotes the Poisson bracket with corresponding Hamiltonian
of the averaged hierarchy. To keep the first multiplier in
(\ref{ave}) finite one has to pass to the slow times $T_{2k-1}$.
So, the observables for the averaged hierarchy
are generated by
\debut
\vev{\vev{\CL _0(T,y\vert B)}}=
\exp (\sum_{k\geq 1} T_{2k-1}\widehat{I}_{2k-1})
\cdot\vev{\langle\ \exp (\sum_{k\geq 1} y_{2k}J_{2k}) \ \rangle}
\non
\fin
Let us repeat that the average in this formulae depends only
on $B_j$ which are supposed to be functions of slow times,
the multiplier with $\widehat{I}_{2k-1}$ corresponds to
derivations with respect to the slow times.
\def\apsi{\vev{\langle \psi ^*\rangle} }

The averaged formulae becomes really beautiful with
the fermionic generating function:
\debut
\vev{\vev{\widehat \CL _0(B)}}\ket{-1}=
 \Delta ^{-1}
\  g(B)
\ \apsi _1\cdots\apsi _n |-1-2n\rangle
\label{aob}
\fin
where
$$\apsi _j=\int\limits _{a_j}
{dAA^{2n-1}\over\sqrt{\CP(A^2)}}\psi ^*(A) , $$
$g(B)$ is the same as in quantum case (\ref{gB}).

Remind that we have rewritten the equations of KdV
hierarchy in the weak sense as follows
$$\langle \Psi _{-5}|\ \CC _{cl}\w= 0,\qquad
\langle \Psi _{-3}|\ \CQ _{cl}\w= 0 $$
For the averaged hierarchy these equations have to
be replaced by
\debut
\langle \Psi _{-5}|\ \CC _0\w= 0,\qquad
\langle \Psi _{-3}|\ \CQ _0\w= 0
\label{weq}
\fin
In this section, the symbol $\w= 0$ means vanishing of the matrix
elements with the averaged generating function (\ref{aob}).
The averaged $\CC _0$ and $\CQ _0$ are
\debut
\CC _0=\int\psi(D){d\over dD}\psi(D) dD,\qquad
\CQ _0=\int\psi(D)d\widehat{I}(D)
\non
\fin
where the terms with $I(D)$ are omitted in the definition of $\CC _0$
comparing with $\CC _{cl}$ (\ref{ccl}) because $dI(D)$ is of order
$\epsilon$, the definition of $\CQ _0$ is practically the
same as in (\ref{qcl}) because the latter is homogeneous in $dI(D)$.
Let us explain the implications of the equations
(\ref{weq}) for the averaged hierarchy.

\proclaim Proposition 15.
 The equation
$$\langle \Psi _{-5}|\CC _0 \w= 0 $$
for the averaged hierarchy follows from the Riemann bilinear relation
for hyper-elliptic integrals.
\par
\proof
 The Riemann bilinear relation for the hyper-elliptic
integrals is equivalent to the formula
$$\int\limits _{c_1}{dA_1\over\sqrt{\CP (A_1^2)}}
\int\limits _{c_2}{dA_2\over\sqrt{\CP (A_2^2)}}\ C(A_1,A_2)=c_1\circ c_2 $$
where $c_1\circ c_2$ is the intersection number of cycles, the
anti-symmetric polynomial $C(A_1,A_2)$ is given by
$$C(A_1,A_2)= C'(A_1,A_2)-C'(A_2,A_1)$$
where
$$C'(A_1,A_2)=
\sqrt{\CP (A_1^2)}{d\over dA_1}\( \sqrt{\CP (A_1^2)} {A_1\over A_1^2-A_2^2}\)
$$
(see, for example, \cite{sm3} for a relevant discussion).

On the other hand we have
$$\CC _0\ g(B)=
g(B)\    \widehat{\CC} _0
$$
where
$$\widehat{\CC} _0 =\int \CP (D^2)D^{-4n}\psi (D){d\over dD}\psi (D) dD $$
Consider the formula for averaged observables
(\ref{aob}). Suppose that we undo the averaging considering
instead of
$$\langle \Psi _{-1}|\apsi _1\cdots\apsi _n |-2n-1\rangle$$
the polynomial
$$ \langle \Psi _{-1}|\psi ^*(A_1)\cdots\psi ^*(A_n) |-2n-1\rangle
\prod_i A_i^{2n-1}$$
The averaging corresponds to
integrating over closed cycles on the surface, so
the latter polynomial is defined up to exact forms
of the following kind
$$\sum\limits _{i=1}^n (-1)^i\ M(A_1^2,\cdots,\widehat{A_i^2},\cdots,A_n^2 )
\ \sqrt{\CP (A_i^2)} {d\over dA_i}\( \sqrt{\CP (A_i^2)} Q(A_i)\)$$
where $Q(A)$ is an arbitrary polynomial and
$ M(A_1^2,\cdots,\cdots,A_{n-1}^2 ) $ is anti-symmetric.

It is clear that the statement of the proposition will be proven
if we show that for every $\langle \Psi _{-5}|$
up to exact forms one has
\debut
&&\langle \Psi _{-5}|\widehat{\CC} _0 \psi ^*(A_1)\cdots\psi ^*(A_n)
|-1-2n\rangle \prod_j A_j^{2n-1}\simeq\non\\
&& \simeq\sum\limits _{i<j}(-1)^{i+j}M(A_1^2,\cdots ,\widehat{A_i^2},
\cdots ,\widehat{A_j^2},\cdots A_n^2)C (A_i,A_j)
\label{ccccc}
\fin
for some anti-symmetric $ M(A_1^2,\cdots ,A_{n-2}^2 )$.

Just like in the proof of Proposition 2 from Section 3
we have three possibilities for $\langle \Psi _{-5}| $ which can
give a non-trivial result:

1. The depth of $\langle \Psi _{-5}|$ is greater than $-2n-1$

2. The vector $\langle \Psi _{-5}|$
is obtained from a vector $\langle \Psi _{-1}|$
whose depth is greater than $-2n-1$ by application

of $\psi ^{*}_{-2p-1} \psi ^{*}_{-2q-1}$
with $q>p\ge n$ (i.e. there are two holes below $-2n-1$).

3.The vector
$\langle \Psi _{-5}|$
is obtained from a vector $\langle \Psi _{-3}|$
whose depth is greater than $-2n-1$ by application

of $\psi ^{*}_{-2p-1}$
with $p\ge n$ (i.e. there is one hole below $-2n-1$).
\newline
In the first case using the identity
$$2\int\limits _{|D|>|A_1|,|A_2|} {\sqrt{\CP(D^2)}\over D^2-A_1^2}
{d\over dD}\({\sqrt{\CP(D^2)}\over D^2-A_2^2}\)DdD= C(A_1,A_2)$$
one easily gets the formula
(\ref{ccccc}) with
$$ M(A_1^2,\cdots ,A_{n-2}^2 ) =
\langle \Psi _{-5}|\psi ^*(A_1)\cdots\psi ^*(A_{n-2})
|-1-2n\rangle \prod_j A_j^{2n-1}
$$
In the second case it is necessary
that in the expression
$ \langle \Psi _{-1} | \psi ^{*}_{-2p-1}
\psi ^{*}_{-2q-1} \ \widehat{\CC} _0 $
the operator $\widehat{\CC} _0$ annihilates two holes.
Hence one find the integral
$$
\langle \Psi _{-1} |
2(p-q)\int \CP (D^2)D^{-4n+2p+2q+1} dD =0
$$
It vanishes because $p,q\ge n$. \newline
Finally, in the third case
it is necessary that in the  expression
$ \langle \Psi _{-3} | \psi ^{*}_{-2p-1}
\ \widehat{\CC}_0$
the operator $ \widehat{\CC}_0$ annihilates the hole. This gives
$$\langle \Psi _{-3} |\int\(\CP(D^2)+D{d\over dD}\CP(D^2)\)D^{-4n+2p}\psi (D)dD
$$
So, in the matrix elements we shall find the polynomials
$$\int\limits _{|D|>|A_j|}
\(\CP(D^2)+D{d\over dD}\CP(D^2)\)D^{-4n+2p}{1\over D^2-A_j^2}dD =
2\sqrt{\CP (A_j^2)} {d\over dA_j}\(\sqrt{\CP (A_j^2)}A_j^{2p-2n+1}\)$$
which corresponds to an exact form. \square

The similarity of the proof of this proposition with that of
Proposition 2 of Section 3 is quite impressive.

Let us consider now the operator $\CQ _0$. It is responsible for the
equation of motion as shows the following proposition.

\proclaim Proposition 16. The equations
\debut
{\partial \over \partial{T_{2p+1}}}d\tilde{\omega} _{2q+1}(A)=
{\partial \over \partial{T_{2q+1}}}d\tilde{\omega} _{2p+1}(A)
\label{UU}
\fin
follow from
\debut
\langle \Psi _{-3}|\CQ _0\w= 0\qquad \langle \Psi _{-5}|\CC _0\w= 0
\label{WWW}
\fin
\par
\proof
We will show that the Whitham equations (\ref{UU}) follow by considering the vectors
$$\langle \Psi _{-3}| =\langle -1|\psi ^*_{-2p-1}\psi ^*_{-2q-1}\psi _{2s+1}$$
The proof goes in two steps. First we shall show that the equation (\ref{WWW})
implies that:
\debut
(2p+1)\widehat{I}_{2q+1}\langle -1|\psi _{2p+1}\psi ^* (A)
\ -\ (2q+1)\widehat{I}_{2p+1}\langle -1|\psi _{2q+1}\psi ^* (A) \w= 0
\label{eqsup}
\fin
Indeed, applying  $\CQ _0$ to this vector $\langle \Psi _{-3}|$ gives
$$\langle -1|\psi ^*_{-2p-1}\psi ^*_{-2q-1}\psi _{2s+1}\ \CQ _0 =
\widehat{I}_{2p+1} \langle -1|\psi ^*_{-2q-1}\psi _{2s+1}
\ -\ \widehat{I}_{2q+1} \langle -1|\psi ^*_{-2p-1}\psi _{2s+1}
\w= 0
$$
Now notice that
$$(2s+1)\langle -1|\psi ^*_{-2p-1}\psi _{2s+1}=
(2p+1)\langle -1|\psi ^*_{-2s-1}\psi _{2p+1}\ +
\ \langle -1|\psi ^*_{-2s-1}\psi ^*_{-2p-1} \CC _0 $$
Hence having in mind the equation $\langle \Psi _{-5}|\ \CC _0\w= 0$
one gets
$$ (2p+1)\widehat{I}_{2q+1}\langle -1|\psi _{2p+1}\psi ^*_{-2s-1}\ -
\ (2q+1)\widehat{I}_{2p+1}\langle -1|\psi _{2q+1}\psi ^*_{-2s-1} \w= 0
$$
Since it is true for every $s$ we can write it for the generating function
as in the eq.(\ref{eqsup}).

The second step consists in computing the following average
$$\Delta ^{-1}\langle -1|\psi _{2p+1}\psi ^* (A)
\ g(B)
\ \apsi _1\cdots \apsi _n|-1-2n\rangle $$
Noticing that
$$ \langle -1|\psi _{2p+1}\psi ^* (A)\ g(B)
= \langle -1|
\int dD D^{-2n+2p}\sqrt{\CP(D^2)}
{A^{2n}\over \sqrt{\CP(A^2)}}\psi (D)\psi ^*(A) $$
and calculating the matrix element in a usual way we get the answer:
\debut
&&\Delta ^{-1}\langle -1|\psi _{2p+1}
\psi ^* (A)\ g(B)
\ \apsi _1\cdots \apsi _n|-1-2n\rangle  =
\Delta ^{-1}A\ det (M(A))
\non\fin
where $M(A)$ is $(n+1)\times (n+1)$ matrix with the following matrix
elements
\debut
&&M(A)_{i,j}=\int\limits _{a_j}{D^{2(i-1)}\over \sqrt{\CP(D^2)}} dD,
\quad i,j=1,\cdots ,n \non\\
&&M(A)_{i,n+1}={A^{2(i-1)}\over \sqrt{\CP(A^2)}}, \quad i=1,\cdots ,n \non\\
&&M(A)_{n+1,j}= \int\limits _{a_j}{Q_p(D^2)\over \sqrt{\CP(D^2)}},
\quad j=1,\cdots ,n \non\\
&&M(A)_{n+1,n+1}= {Q_p(A^2)\over \sqrt{\CP(A^2)}}
\non
\fin
where
$$
Q_p(A^2)=\int\limits _{|D|>|A|}
dD {D^{2p+1}\over D^2-A^2} \sqrt{\CP(D^2)}=
\[ \sqrt{\CP(A^2)}A^{2p}\]_+
$$
where $[\cdots]_+$ means taking the polynomial part in the
expansion around infinity.
It is quite obvious that the normalized differential  $d\tilde{\omega} _{2p+1}(A)$
is given by
$$d\tilde{\omega} _{2p+1}(A) =(2p+1) \Delta ^{-1}\ det (M(A)) dA $$
which finishes the proof of the proposition.
Returning to the beginning of the proof one find that the
expression
$$ (2p+1)\langle -1|\psi _ {2p+1}\psi ^*(A){dA\over A}$$
can be considered as "symbol" of the normalized differential
$d\tilde{\omega} _{2p+1}(A)$. \square

The equation $\langle \Psi _{-3}|\ \CQ _0\w= 0$ for more complicated
states $\langle \Psi _{-3}| $
than those considered in the Proposition 16
implies other linear partial differential
equations for $B_j$, so, we get the whole Whitham hierarchy. However
the equations (\ref{UU}) are the only ones with derivatives with
respect to only two times. We shall not go further into the study
of the Whitham hierarchy, because it is not our goal. What we really wanted
to do was to show the remarkable parallel between the Whitham method and
the quantum form factor formulae. We hope that this goal is achieved.

{\bf Acknowledgement} We would like to thank Tetsuji Miwa for
his interest in this work and careful reading of the manuscript.

\section{Appendix A}
In this appendix we explain why the conditions
\debut
&&\left . L_\CO^{(n)} (A_1,\cdots ,A_n|B_1,\cdots ,B_{2n})
\right|_{B_{2n}=-B_{1},\ A_n=\pm B_{2n-1}}= \non\\
&&\hskip 1cm
=-\epsilon ^{\pm}L_\CO^{(n-1)} (A_1,\cdots ,A_{n-1}|B_2,\cdots ,B_{2n-1})
\label{ress}
\fin
($\epsilon =+$ or $-$ respectively for the operators $\Phi _{2k} $
and their descendents or  $\Phi _{2k+1} $
and their descendents) 
and 
\debut res _{A_n=\infty}\(
\prod\limits _{i=1}^n \prod\limits _{j=1} ^{2n} \psi (A_i,B_j)
\prod\limits _{i<j} (A_i^2-A_j^2)
\ L_\CO^{(n)} (A_1,\cdots ,A_n|B_1,\cdots ,B_{2n})a _n^{k}\)=0,
\quad k\ge n+1
\label{vvv}\fin
are sufficient for the locality
of the operator whose form factors are given by
\debut
&&f_\CO (\b _1,\b _2,\cdots ,\b _{2n})_{- \cdots -+\cdots +}
=\non \\&&
\hskip -3cm
=c^n \prod\limits _{i<j}\zeta (\b _i-\b _j)
\prod\limits _{i=1}^n \prod\limits _{j=n+1} ^{2n}
{1\over\sinh{\nu}(\b _j-\b _i-\pi i)}
\exp(-{1\over 2}(\nu (n-1)-n)\sum \b _j)
\non \\ &&\times
\widehat{f}_\CO (\b _1,\b _2,\cdots ,\b _{2n})_{-\cdots -+\cdots +}
\non
\fin
where
\debut
\widehat{f}_\CO (\b _1,\b _2,\cdots ,\b _{2n})_{-\cdots -+\cdots +}&=&
\non\\
&&\hskip -5cm
={1\over (2\pi i)^n}\int dA_1\cdots \int dA_n
\prod\limits _{i=1}^n \prod\limits _{j=1} ^{2n} \psi (A_i,B_j)
\prod\limits _{i<j} (A_i^2-A_j^2)
\ L_\CO^{(n)} (A_1,\cdots ,A_n|B_1,\cdots ,B_{2n})
\prod\limits _{i=1}^n a_i^{-i}
\non\fin
The calculations which we are going to make are well known even for
more general case \cite{book}. However, we want to repeat them
for our particular situation for the completeness of the exposition.

In the case of diagonal scattering the only non-trivial
requirement for the form factors is the following:
\debut
&&2\pi i\ res_{\b _{2n}=\b _{1}+\pi i}
f_\CO (\b _1,\b _{2},\cdots ,\b _{2n-1},\cdots \b _{2n})
_{-\cdots -+\cdots +}=\non\\&&\hskip 1cm=
f_\CO (\b _2,\cdots ,\cdots ,\b _{2n-1})_{-\cdots -+\cdots +}
\left(\prod\limits _{j=2}^{2n-1}S(\b _j-\b _1)
-\epsilon \right)
\label{loca} \fin
where the S-matrix in our case is
$$S(\b _2-\b_1)=
\prod\limits _{j=1}^{\nu -1}{\sinh{1\over 2}(\b _2-\b_1 +{\pi i\over\nu}j)
\over\sinh{1\over 2}(\b _2-\b_1 -{\pi i\over\nu}j)}=
{\psi (-B_1,B_2)\over \psi (B_1,B_2)}$$
Using the identity \cite{book}
$$\exp (-{1\over 2}(\nu -1)(\b _1+\b _j))
{\zeta (\b _1-\b _j)\zeta (\b _j-\b _1-\pi i)
\over (\sinh \nu(\b _j-\b _1))^2}= {1\over \psi (B_1,B_j)}  $$
one find that the relation (\ref{loca}) is equivalent to
\debut
\left.\widehat{f}_\CO (\b _1,\b _2,\cdots ,\b _{2n-1},
\b _{2n})_{-\cdots -+\cdots +}\right|_ {\b _{2n}=\b _{1}+\pi i}
=\non \\
={1\over 2B_1}\widehat{f}_\CO (\b _2,\cdots ,\b _{2n-1})_{-\cdots -+\cdots +}
\left(\prod\limits _{j=2}^{2n-1}\psi (-B_1,B_j)
-\epsilon\prod\limits _{j=2}^{2n-1}\psi (B_1,B_j)\right)
\non
\fin
if the constant $c$ is taken as $c=2\nu (\zeta (-\pi i))^{-1}$.
Explicitly the LHS of this equation is
\debut
&&{1\over (2\pi i)^n}\int dA_1\cdots \int dA_n
\prod\limits _{i=1}^n
{a_i-b_1\over A_i^2-B_1^2}
\prod\limits _{j=2} ^{2n-1} \psi (A_i,B_j)
\non\\&&\hskip 1.5cm
\times\prod\limits _{i<j} (A_i^2-A_j^2)
\ L_\CO^{(n)} (A_1,\cdots ,A_n|B_1,\cdots ,B_{2n-1},-B_1)
\prod\limits _{i=1}^n a_i^{-i} \label{innt}
\fin
where we have used the identity
$$\psi(A,B)\psi(A,-B)={a-b\over A^2-B^2} $$
Let us consider the integral over $A_n$. If the contour is such that
$|A_n|>|B_1|$ we can replace in this integral $a_n^{-n}$ by
$b_1^{-n+1}(a_n-b_1)^{-1}$. Indeed
$${b_1^{-n+1}\over(a_n-b_1)} =a_n^{-n} +
\sum\limits _{j=1}^{n-1}a_n^{-j} b_1^{-n+j}+
\sum\limits _{j\ge n+1}a_n^{-j} b_1^{-n+j} $$
the sum $ \sum\limits _{j=1}^{n-1}$ can be omitted due to anti-symmetry
with respect to $a_j$, $j=1,\cdots ,n-1$, the sum
$\sum\limits _{j\ge n+1}$ can be omitted
because the integrand does not have residue at $A_n=\infty$
due to (\ref{vvv}).
Thus the integral over $A_n$ becomes
\debut
&&b_1^{-n+1}{1\over 2\pi i}\int\limits _{A_n>B_1} dA_n
{1\over A_n^2-B_1^2}
\prod\limits _{j=2} ^{2n-1} \psi (A_n,B_j)
\prod\limits _{i<n} (A_i^2-A_n^2)
\ L_\CO^{(n)} (A_1,\cdots ,A_n|B_1,\cdots ,B_{2n-1},-B_1)
=\non \\
&&={1\over 2}b_1^{-n+1} B_1^{-1} \prod\limits _{i<n} (A_i^2-B_1^2)
\left(\prod\limits _{j=2} ^{2n-1} \psi (-B_1,B_j)
L_\CO^{(n)} (A_1,\cdots ,A_{n-1},-B_1|B_1,\cdots ,B_{2n-1},-B_1) -\right.\non\\
&&-\left.\prod\limits _{j=2} ^{2n-1} \psi (B_1,B_j)
L_\CO^{(n)} (A_1,\cdots ,A_{n-1},B_1|B_1,\cdots ,B_{2n-1},-B_1) \right)
\non
\fin
Let us substitute this expression into (\ref{innt}):
\debut
&&{1\over 2B_1}{1\over (2\pi i)^{n-1}}
\int dA_1\cdots \int dA_{n-1}
\prod\limits _{i=1}^{n-1}
\prod\limits _{j=2} ^{2n-1} \psi (A_i,B_j)
\prod\limits _{i<j} (A_i^2-A_j^2)
b_1^{-n+1}\prod\limits _{i=1}^{n-1} a_i^{-i}(a_i-b_1) \non\\ &&\times
\left(\prod\limits _{j=2} ^{2n-1} \psi (-B_1,B_j)
L_\CO^{(n)} (A_1,\cdots ,A_{n-1},-B_1|B_1,\cdots ,B_{2n-1},-B_1) -
\right.\non\\
&&-\left.\prod\limits _{j=2} ^{2n-1} \psi (B_1,B_j)
L_\CO^{(n)} (A_1,\cdots ,A_{n-1},B_1|B_1,\cdots ,B_{2n-1},-B_1) \right)
=\non \\=
&&{1\over 2B_1}{1\over (2\pi i)^{n-1}}
\int dA_1\cdots \int dA_{n-1}
\prod\limits _{i=1}^{n-1}
\prod\limits _{j=2} ^{2n-1} \psi (A_i,B_j)
\prod\limits _{i<j} (A_i^2-A_j^2)
\prod\limits _{i=1}^{n-1} a_i^{-i} \non\\ &&\times
\left(\prod\limits _{j=2} ^{2n-1} \psi (-B_1,B_j)
L_\CO^{(n)} (A_1,\cdots ,A_{n-1},-B_1|B_1,\cdots ,B_{2n-1},-B_1) -
\right.\non\\&&-\left.\prod\limits _{j=2} ^{2n-1} \psi (B_1,B_j)
L_\CO^{(n)} (A_1,\cdots ,A_{n-1},B_1|B_1,\cdots ,B_{2n-1},-B_1) \right)
\label{integr}
\fin
where we have replaced $b_1^{-1}a_i^{-i}(a_i-b_1)$ by $-a_i^{-i}$
for the following reason: for $a_1$ this expression is $ b_1^{-1} -a_1^{-1} $,
the integrand with $ b_1^{-1}$ is regular at zero, for $a_i$ with
$i>1$ we use anti-symmetry.
The final formula (\ref{integr}) shows that the equation (\ref{ress})
is sufficient for locality.

\section{Appendix B}

The reflectionless case is a rather degenerate one, so, the
deformed Riemann bilinear identity \cite{sm2} does not exist
in complete form. However for our needs we use only the
consequence of the deformed Riemann bilinear identity which
allows a simple proof in the reflectionless case. We give
this proof here for the sake of completeness.
We have used the following fact:
\debut
\int dA_1 \int dA_2
\prod\limits _{i=1}^2 \prod\limits _{j=1} ^{2n} \psi (A_i,B_j)
C(A_1,A_2 ) a_1^{k}a_2^{l}=0\quad \forall k,l \label{ii}
\fin
where we had the expression for $C(A_1,A_2)$:
\debut
C(A_1,A_2 )={1\over A_1A_2}\left\{ {A_1-A_2\over A_1+A_2 }
(P(A_1)P(A_2)-P(-A_1)P(-A_2))
+
(P(-A_1)P(A_2)-P(A_1)P(-A_2))\right\}
\label{ccc} \fin
Let us introduce the functions
$$F(A)=\prod\limits _{j=1} ^{2n} \psi (A,B_j)P(A),
\qquad G(A)= \prod\limits _{j=1} ^{2n} \psi (A,B_j)P(-A) $$
Recall that the function $\psi(A,B)$ satisfies the difference
equation
\debut
\psi(Aq,B)=\left({B-A\over B+qA}\right)~ \psi(A,B),
\label{fonc} \fin
which implies that
$$F(Aq)=G(A) $$
The integral (\ref{ii}) can be rewritten as follows:
\debut
\int {dA_1\over A_1} \int {dA_2 \over A_2}
\left\{ {A_1-A_2\over A_1+A_2 }
(F(A_1)F(A_2)-F(qA_1)F(qA_2))+
(F(qA_1)F(A_2)-F(A_1)F(qA_2))\right\}
a_1^{k}a_2^{l}  \non
\fin
Changing variables $A_i\to qA_i$ where needed one easily find
that this integral equals zero. Recall that $a_i=A_i^{2\nu}$ and
$q^{2\nu}=1$, so $ a_1^{k}a_2^{l} $ do not change under these changes of
variables.

Similar proof for the case of generic coupling constant is more
complicated because one crosses singularities when changing
variables and moving contours.

Now let us prove the Proposition 1.
We want to find equivalent expressions for $C(A_1,A_2)$.
Let us rewrite the expression (\ref{ccc}) in the integral form:
$$C(A_1,A_2)= C'(A_1,A_2)- C'(A_2,A_1) $$
where
\debut
&&C'(A_1,A_2)=
{1\over 2A_2}\left( {P(A_1)P(A_2)\over A_2+A_1}
+{P(A_1)P(-A_2)\over A_2-A_1}
-{P(-A_1)P(A_2)\over A_2-A_1}
-{P(-A_2)P(-A_1)\over A_2+A_1} \right)=
\non\\ &&=A_1
\int\limits _{|D_2|>|D_1|}dD_2 \int\limits _{|D_1|>|A_1|,|A_2|}dD_1
{P(D_1)P(D_2)\over (D_1+D_2)(D_1^2-A_1^2)(D_2^2-A_2^2)}\non\label{jj}
\fin
Let us modify this expression by adding "exact forms" in variables $A_1$.
It is convenient to use the formula
\debut
&&Q(A_1)P(A_1)-qQ(qA_1)P(-A_1)=\non\\
&&\hskip 0.5cm=\int\limits _{|D_1|>|A_1|}dD_1
{P(D_1)\over (D_1^2-A_1^2)} (Q(D_1)(D_1+A_1) -qQ(-qD_1)(D_1-A_1))
\non\fin
Suppose that the polynomial $Q(A)$ solves the equation
\debut
Q(D_1)+qQ(-qD_1)=
\int\limits _{|D_2|>|D_1|}dD_2
{P(D_2)\over (D_1+D_2)(D_2^2-A_2^2)}\label{eqQ}
\fin
(obviously the RHS of this equation is a polynomial) then we
can rewrite the expression for $C'(A_1,A_2)$ in the following equivalent
form
$$C'(A_1,A_2)=
\int\limits _{|D_1|>|A_1|}dD_1D_1
{P(D_1)\over (D_1^2-A_1^2)} (Q(D_1)-qQ(-qD_1)) $$
Now we have to solve the equation (\ref{eqQ}). It is simple:
$$
Q(D_1)={1\over D_1}
\int\limits _{|D_2|>|D_1|}dD_2 \eta \left({D_1\over D_2}\right)
{P(D_2)\over (D_2^2-A_2^2)}
$$
where $\eta(x)$ satisfy $\eta(x)-\eta(-qx)=\frac{x}{1+x}$. Namely,
$$\eta (x)=\sum\limits_{k=0}^{\infty}{1\over 1+q^{2k+1}}x^{2k} -
\sum\limits_{k=1}^{\infty}{1\over 1-q^{2k}}x^{2k-1} $$
Hence
$$C'(A_1,A_2)=\int\limits _{|D_2|>|D_1|}dD_2
\int\limits _{|D_1|>|A_1|}dD_1 P(D_1)   P(D_2)\tau _e\(D_1\over D_2\)
{1\over (D_1^2-A_1^2)(D_2^2-A_2^2)}
$$
where $\tau_e(x)=\eta(x)+\eta(-qx)$, ie.
$$\tau _e(x)=\sum\limits_{k=1}^{\infty}{1-q^{2k-1}\over 1+q^{2k-1}}x^{2k-1} -
\sum\limits_{k=1}^{\infty}{1+q^{2k}\over 1-q^{2k}}x^{2k} $$
The expression for $C_e(A_1,A_2)$ given in Proposition 1
follows from these formulae. The expression for $C_o(A_1,A_2)$ can be
obtained in a similar way eliminating the even degrees of $A_1$ from
$C'(A_1,A_2)$.

\end{document}